\documentclass[pra,twocolumn,a4paper,superscriptaddress,showkeys]{revtex4-1}

\usepackage{amsmath}
\usepackage{mathptmx}

\usepackage{color}
\usepackage{subfigure}
\usepackage{paralist}
\usepackage{appendix}
\usepackage[normalem]{ulem}
\usepackage{mathtools}

\usepackage {graphicx}
\graphicspath{{./figs/}}

\newcommand {\qfig}[1]{Fig.~\ref{#1}}
\newcommand {\qsect}[1]{Sect.~\ref{#1}}

\newcommand {\queq}[1]{(\ref{#1})}
\newcommand {\qeq}[1]{Eq.~\queq{#1}}

\newcommand {\beql}[1]{\begin{equation} \label{#1}}
\newcommand {\eeql}{\end{equation}}
\newcommand {\beq}{\begin{equation}}
\newcommand {\eeq}{\end{equation}}

\newcommand {\etal}{\emph{et al.}\ }

\begin{document}

\title{%
%Nanoindentation in metallic glasses \\
Nanoscratching of metallic glasses -- An atomistic study
}

\author{Karina E. Avila}
\affiliation{%
Physics Department,
University Kaiserslautern,
Erwin-Schr{\"o}dinger-Stra{\ss}e, D-67663 Kaiserslautern, Germany}
\affiliation{%
Research Center OPTIMAS,
University Kaiserslautern,
Erwin-Schr{\"o}dinger-Stra{\ss}e, D-67663 Kaiserslautern, Germany}

\author{Stefan K\"uchemann}
\affiliation{%
Physics Department,
University Kaiserslautern,
Erwin-Schr{\"o}dinger-Stra{\ss}e, D-67663 Kaiserslautern, Germany}

\author{Iyad Alabd Alhafez}
\affiliation{%
Physics Department,
University Kaiserslautern,
Erwin-Schr{\"o}dinger-Stra{\ss}e, D-67663 Kaiserslautern, Germany}
\affiliation{%
Research Center OPTIMAS,
University Kaiserslautern,
Erwin-Schr{\"o}dinger-Stra{\ss}e, D-67663 Kaiserslautern, Germany}

\author{Herbert M.~Urbassek}
\email{urbassek@rhrk.uni-kl.de}
\homepage{http://www.physik.uni-kl.de/urbassek/}
\affiliation{%
Physics Department,
University Kaiserslautern,
Erwin-Schr{\"o}dinger-Stra{\ss}e, D-67663 Kaiserslautern, Germany}
\affiliation{%
Research Center OPTIMAS,
University Kaiserslautern,
Erwin-Schr{\"o}dinger-Stra{\ss}e, D-67663 Kaiserslautern, Germany}

\date{\today}

\begin{abstract}

Tribological properties of materials play an important role in engineering applications. Up to now, a number of experimental studies have identified correlations between tribological parameters and the mechanical response. Using molecular dynamics simulations, we study abrasive wear behavior via nanoscratching of a Cu$_{64.5}$Zr$_{35.5}$ metallic glass. The evolution of the normal and transverse  forces and  hardness values follows the behavior well known for crystalline substrates. In particular, the generation of the frontal pileup weakens the response of the material to the scratching tip and leads to a decrease of the transverse hardness as compared to the normal hardness. However, metallic glasses soften with increasing temperature, particularly above the glass transition temperature thus showing a higher tendency to structurally relax an applied stress. This plastic response is analyzed focusing on local regions of atoms which underwent strong von-Mises strains, since these are the basis of shear-transformation zones and shear bands. The volume occupied by these atoms increases with temperature, but large increases are only observed above the glass transition temperature. We quantify the generation of plasticity  by the concept of \emph{plastic efficiency}, which relates the generation of plastic volume inside the sample with the formation of external damage, viz.\ the scratch groove. 
In comparison to nanoindentation, the generation rate of the plastic volume during nanoscratching is significantly temperature dependent making the glass inside more damage-tolerant at lower temperature but more damage-susceptible at elevated temperatures.

\end{abstract}


\maketitle

\section{Introduction}

The deformation behavior of metallic glasses can be distinguished into three different responses: Elastic response, visco-elastic response and plastic response. The elastic strain describes deformation which depends linearly on the applied load or stress below the elastic limit which is about 2 \% in metallic glasses~\cite{TCS*12,JS05}. An elastically deformed system returns almost instantaneously after the stress is removed to the initial shape.

The visco-elastic response constitutes one of the main differences in the mechanical response between glasses and crystals since this deformation mode does not occur in the latter material class. It is associated to creep behavior which implies that the system reacts to an applied force on a certain temperature-dependent timescale which can be described, for instance, by the Kohlrausch-William-Watts function~\cite{WMS96}. The timescale depends on the structural state of the sample and, accordingly, it can be decreased via annealing or enhanced via rejuvenation mechanisms~\cite{KM17,KDL*18}.

The third deformation mode, plastic deformation, describes an irreversible deformation process which can be divided into two categories, homogeneous and heterogeneous plastic deformation~\cite{NW06,Spa77,SLN04}. When the metallic glass is deformed at higher temperatures or with low deformation rates, the deformation rate may reach the timescale of the relaxation rate. In this case the plastic deformation is  homogeneously distributed throughout the entire sample~\cite{HDJS07}. During heterogeneous deformation, the deformation rate is significantly above the relaxation rate so the strain is localized into thin planar regions with a thickness of 10--200 nm, called shear bands~\cite{LRKM17,KLD*18}. These shear bands exhibit a severely deformed core featuring a significantly decreased local density ~\cite{LRKM17}, enhanced enthalpy~\cite{KWS*13} and an enhanced diffusion coefficient~\cite{BDRW11}. Eventually, shear bands lead to the failure of the material, which limits the industrial applications of metallic glasses.

Recently, it has been shown that even when the system is able to completely recover the initial shape, local plastic shear events occur during the deformation. These regions  are called shear transformation zones (STZs), consisting of about $\sim$ 100~atoms~\cite{PISC08,ZSJM06,SKW*11} and they are  characteristics of the plastic deformation of amorphous materials~\cite{PISC08,FL98}.

Even when  industrial applications of bulk metallic glasses are still limited, the combination of the features of deformation modes makes metallic glasses often a preferred candidate for applications of engineering materials, such as coatings, mechanical components like gears or magnetic read/write heads~\cite{GRH02}. For instance, the elastic limit is significantly higher for metallic glasses in comparison to crystals, and due to their ability to localize plastic strain into shear bands they are not as brittle as ceramics. In this context, the tribological properties of metallic glasses attracted research interest in the last decades~\cite{THAA09,HN04,HCS*10,ZLS*08,GRH02}. In comparison to crystals, metallic glasses exhibit advantages under certain circumstances. For instance, some research demonstrated that certain compositions of metallic glasses exhibit a superior resistance~\cite{GRH02} to sliding wear.

The prototypical tribological event on the nanoscale is a scratch, in which a tip is indented into the substrate and then moved at a fixed depth along the surface. Such a process allows to identify several material properties, namely the hardness in normal and transverse direction and the friction coefficient.  Such a process is well suited for study by molecular dynamic (MD) simulation. Indeed, quite an impressive number of nanoscratching simulations have been performed on crystalline substrates; the role of dislocation generation and reactions has been investigated as well as the influence of surface orientation and scratch direction~\cite{JM14,GRU14,GBKU15,LLL*16,GAS*17}. However, only few studies have been devoted up to now to the investigation of nanoscratching of metallic glasses~\cite{W16}.

In this paper, we assess the material properties during abrasive wear of a large amorphous Cu$_{64.5}$Zr$_{35.5}$ system in molecular dynamic simulations. We measure the deformation dependence on temperature and scratching depth. This work aims at correlating the mechanical properties -- such as hardness and plastic activity -- and the tribological properties -- friction, wear volume and depth -- of our system. We also compare the differences and similarities between the underlying atomistic deformation mechanisms during abrasive wear and other deformation modes such as nanoindentation. The generation of plastic rearrangements inside the material is measured for the first time in simulations of scratching in metallic glasses. This allows us to  establish a correlation between the friction coefficient and the plasticity inside the sample.  
We also find that the generation of plastic events is less effective in nanoscratching at lower temperature than in indentation but more effective at temperatures around the glass transition temperature.

\section{Method}

We use the open-source code LAMMPS~\cite{Pli95} to simulate the binary-composition metallic glass Cu$_{64.5}$Zr$_{35.5}$. The sample consists of $N=5,619,712$ atoms contained in a cubic simulation box of edge length $L$. Its size varies from $L=450.01$~\AA\  for the lowest temperature to $457.46$ \AA, for the highest temperature. The atomic interactions are modeled by the embedded-atom-method (EAM) potential developed by Mendelev \etal ~\cite{MSK07}. A crystalline mixture was first heated to a temperature above the melting point, $T=2000$ K, for a time period of 500~ps and then cooled to the final temperature with a quenching rate of 0.01 K/ps to obtain the metallic glass.  We note that we studied previously the effect of the quenching rate on the plastic properties of the glass and found that the quenching rate adopted here is sufficiently slow to obtain reliable results on the glass plasticity~\cite{AKAU19}.
The glass transition temperature $T_g$ for this particular composition and potential is around 1000~K for a quenching rate of 1 K/ps 
~\cite{LJP12}; for the quenching rate adopted here there would be a slight change of $T_g$ toward a lower temperature, but not below 800~K.  Here, we  simulated samples at 5 different temperatures. We included $T=1000$~K, which is at or above the glass transition temperature, to observe the effects in the supercooled-liquid regime. 

During the preparation of the sample, periodic boundary conditions were applied in all directions and an isobaric ensemble (at vanishing pressure) with a Nose-Hoover thermostat was used. Once the final temperature is reached, the system is left to relax for a total time of 200 ps with periodic boundary conditions. Then, to prepare the system for the scratch simulations,
periodic boundary conditions are applied in the lateral directions only, while free boundary conditions are exerted on the top surface, and the system is allowed to relax for an additional time of 300 ps.  10 atomic layers at the bottom of the sample are kept fixed  in order to mimic the immobile bulk of a metallic glass in a real experiment and to avoid center-of-mass translation of the entire sample.

The tip has a spherical form with a radius $R=10$ nm. The purely repulsive force exerted by the tip on the system is given by %Ref.~\cite{KPH98}
\begin{equation}
F(r)=K(r-R)^2,
\end{equation} 
where $r$ is the distance of an atom to the center of the indenter. 
The stiffness constant of the tip has been set to $K=10$ eV/\AA$^3$. %Note that this simple but often-employed approach~\cite{RBGU17} ignores any adhesion or friction forces between tip and substrate. 
During indentation we keep the temperature fixed by using an NVT ensemble. 
Our simulation proceeds in 3 steps: First the tip is indented to its final depth $d$; then it is moved at this fixed depth for a distance of 10 nm; finally it is moved out of the workpiece.
%During the scratch simulation, we use fixed boundaries at the bottom to prevent any rigid-body motion of the sample, and periodic boundary conditions at the sides. In addition, a thermostat keeps the glass at a prescribed temperature. Note that, in contrast to experiment, in simulation usually a depth-controlled rather than a load-controlled indentation and scratch process is performed; indeed, load-controlled simulations require more elaborate algorithms and are rare~\cite{J15}.
 In all cases the tip velocity was chosen as  20 m/s. %, see Appendix \ref{appendix3}. %This velocity is high compared to experimental velocities, except for nanotribology experiments. However, velocities of the order of $\mu$m/s, which might be used in scratch experiments, require 7 orders of magnitude more time for computation, and are hence not accessible to MD simulation. The simulation misses in particular thermal processes, which might occur over longer time scales and are improbable to occur during the ns time scale of the simulation. 
In the main part of the paper, d is fixed to 3 nm; in Sect. \ref{s_depth}, the effect of d on the results is studied.
 
 The open source visualization tool OVITO~\cite{Stu10} is used in our work to analyze and visualize  the atomistic configurations.

\section{Results}

\subsection{Groove geometry} \label{s_pile}
\begin{figure}[h]
  \begin{center}
     \includegraphics[width=0.42\textwidth]{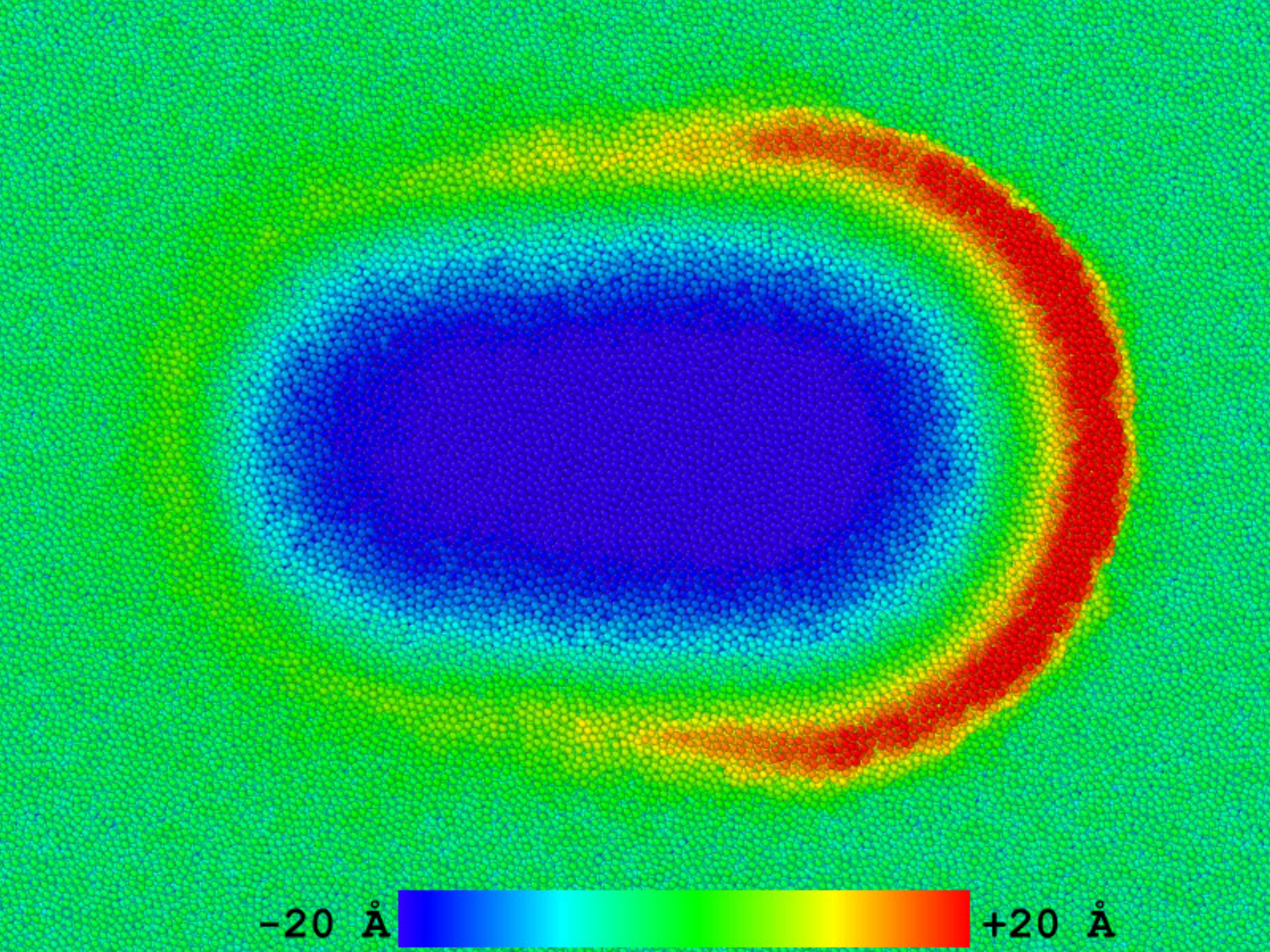}
  \end{center}
  \caption{Pileup generated for temperature $T=10$ K. The  colors represent the height of the generated pileup and the depth of the groove left by the tip.}
  \label{fig:Profile_view}
\end{figure}

We display in  Fig.~\ref{fig:Profile_view} a top view of the scratch groove and the pileup formed above the surface. This image corresponds to a temperature of 10 K; but the grooves and pileups look similar at the other temperatures simulated. Note that both a lateral and a frontal pileup forms. The height of the pileup increases monotonically towards the front of the scratch groove and reaches its highest value immediately in front of the tip.
The pileup also shows a symmetric shape; while this is natural for the glass material scratched here, it is in sharp contrast to the asymmetric pileups generated in crystals~\cite{AU16,GBKU15,ARU18}.

\begin{figure}[h]
  \begin{center}
    \includegraphics[width=0.45\textwidth]{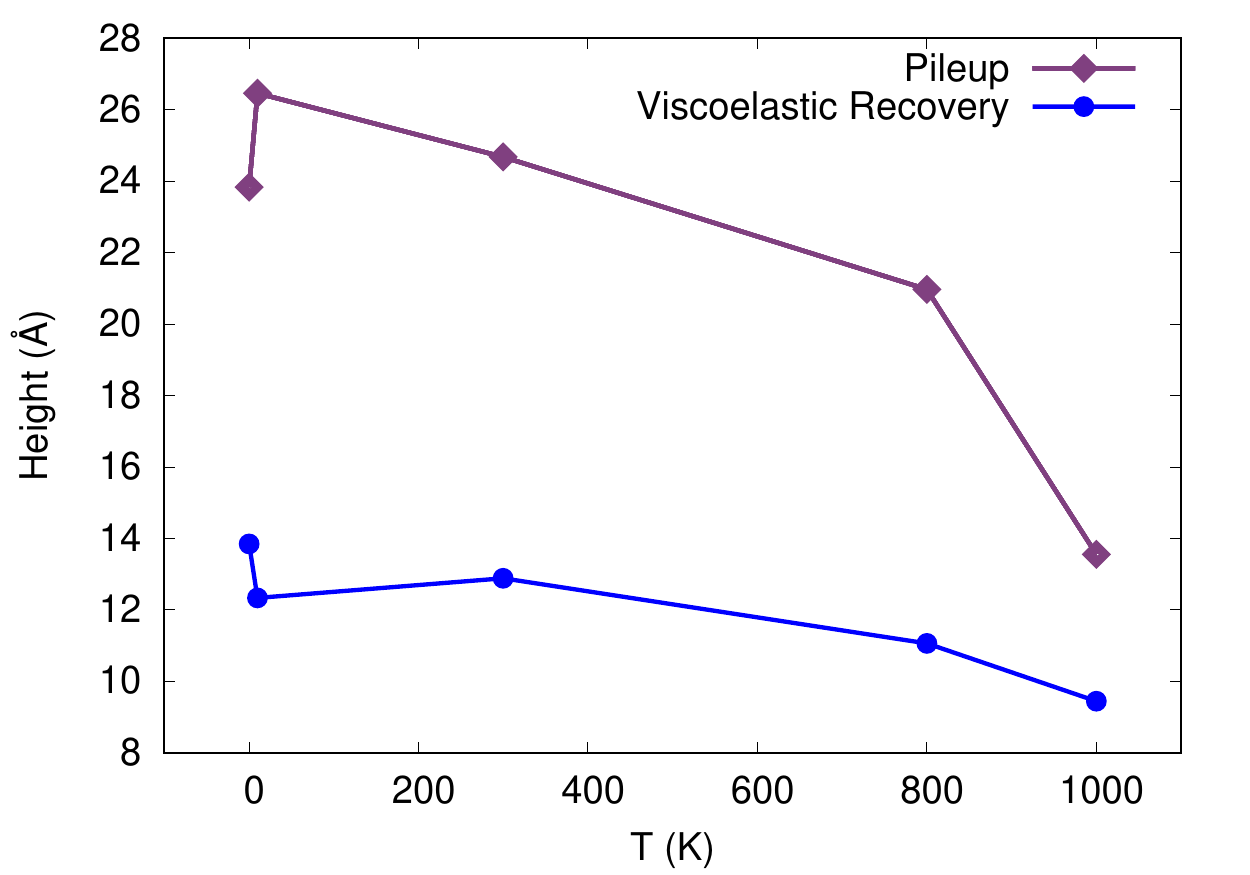}
  \end{center}
  \caption{Pileup and visco-elastic recovery of the groove floor as a function of temperature.
  }
  \label{fig:pileup}
\end{figure}

We determined the height of the pileup above the original surface at its highest point, i.e., immediately in front of the tip. The data were taken after the removal of the indenter and are plotted as a function of temperature in Fig.~\ref{fig:pileup}. The height of the pileup decreases with increasing temperature, in particular at temperatures above the glass temperature. This might be attributed to enhanced temperature--dependent stress relaxation.

Only at the lowest temperature, there is a change in the trend. This might be due to a switch in the atomistic deformation mode to higher shear-band activity which concentrates and mediates the shear strain and effectively directs it to the surface~\cite{HAG*16} but not necessarily to the pileup. Such a switch has indeed been observed in indentation simulations~\cite{AKAU19} and in experiments~\cite{KPA*14}.

While we scratch at a depth of $d=3$ nm, after removal of the tip we find the groove floor to be considerably closer to the initial surface, see Fig.~\ref{fig:pileup}. This phenomenon is well known -- also from  nanoscratching experiments~\cite{HN04,HCS*10} -- and is caused by elastic and visco-elastic 
recovery (or rebound) leading to a decompression of the material below the groove that had been under high compressive stress. 
We measure the rebound of the groove floor at the geometrical center of the scratching groove. The magnitude of the rebound is determined as the height of the groove floor at this point after indenter removal relative to the height of the floor when the indenter was scratching at this position. We find the rebound almost independent of temperature. A slight decrease of the rebound can be observed towards the highest temperatures, which might be due to the elastic response that is known to decrease at high temperatures due to the stress relaxation induced in the sample~\cite{Wan12a}. 

\subsection{Forces, contact areas  and hardness}  \label{s_F}

The forces acting on the indenter tip are reliably determined both in simulation and experiment. We differentiate the normal force, $F_n$, acting in the direction perpendicular to the surface plane and the transverse force, $F_t$, acting along the scratch direction.

\begin{figure*}
  \begin{center}
    \includegraphics[width=0.45\textwidth]{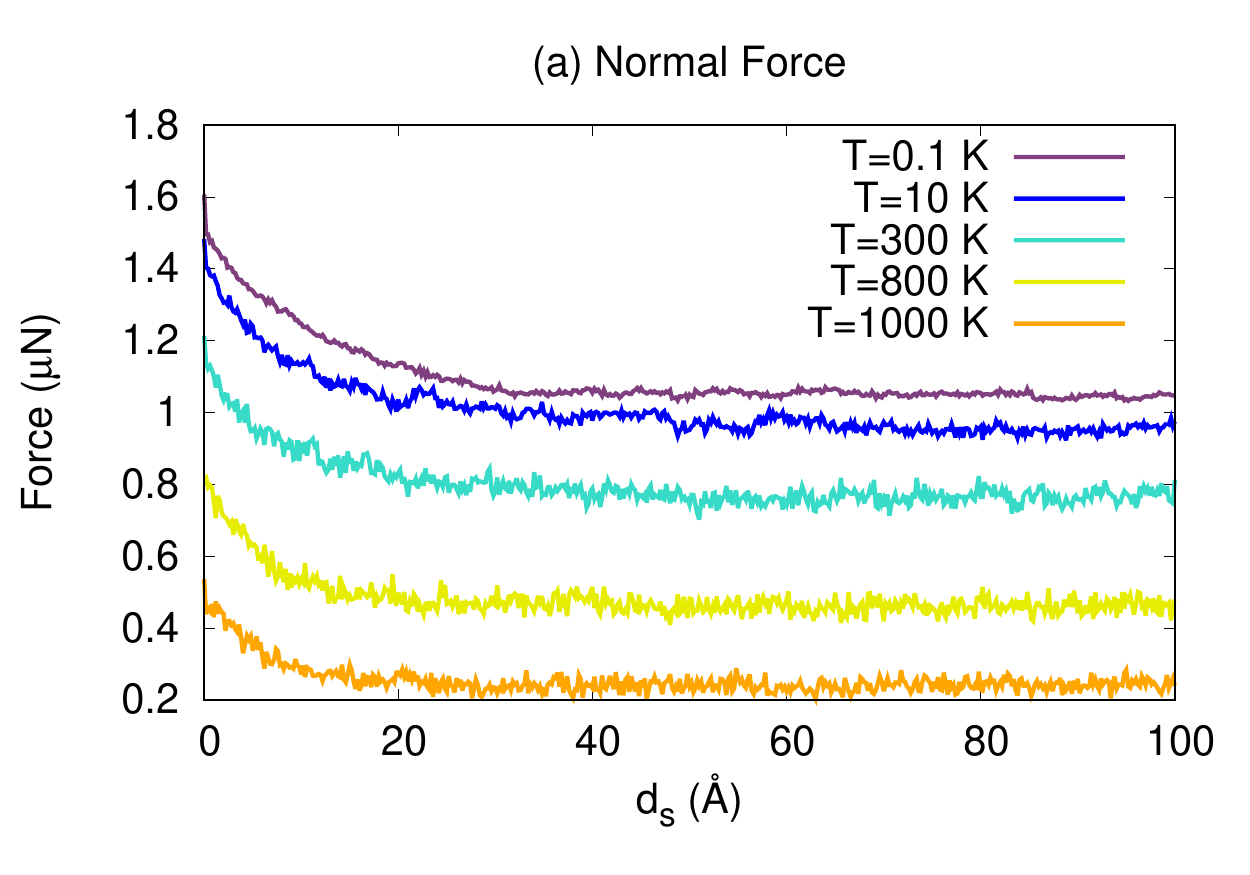}
    \includegraphics[width=0.45\textwidth]{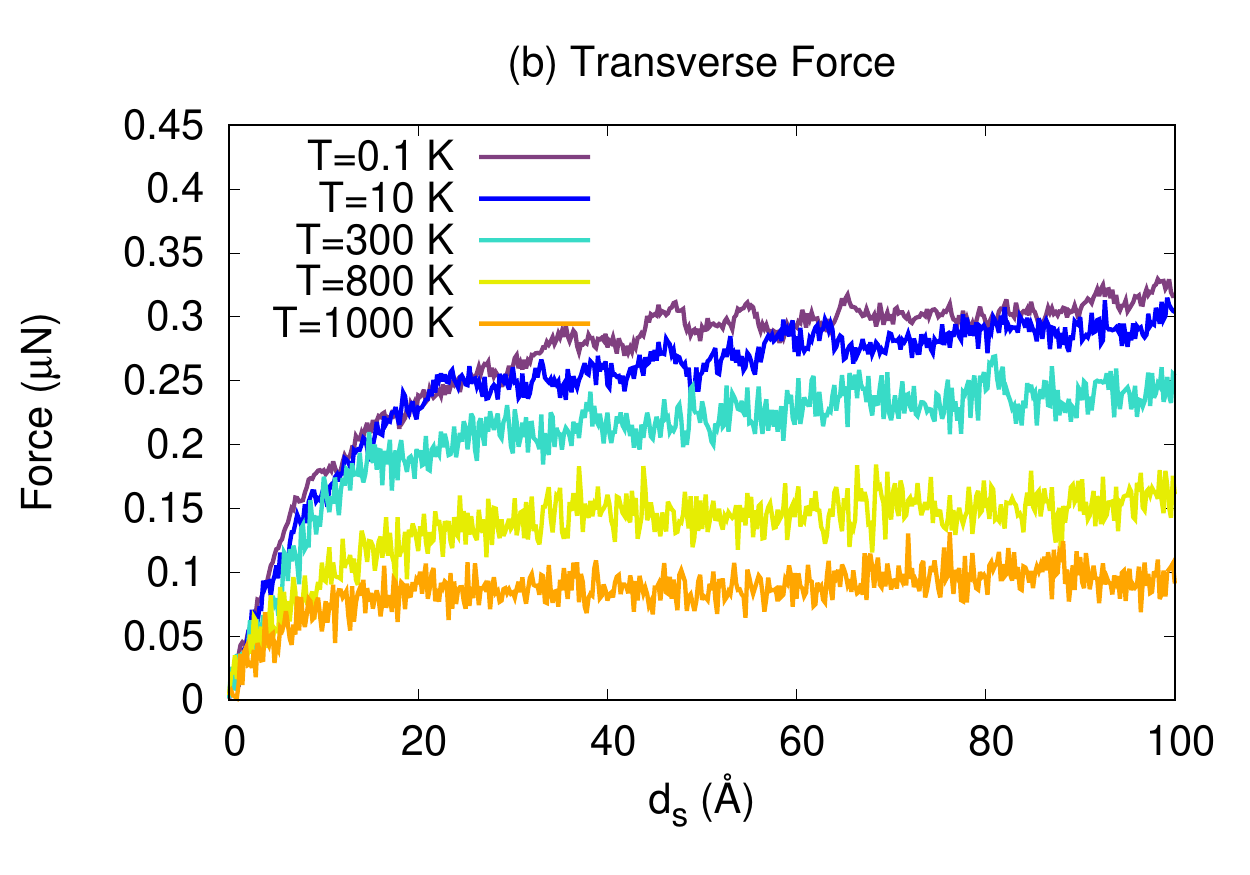}\\
    \includegraphics[width=0.45\textwidth]{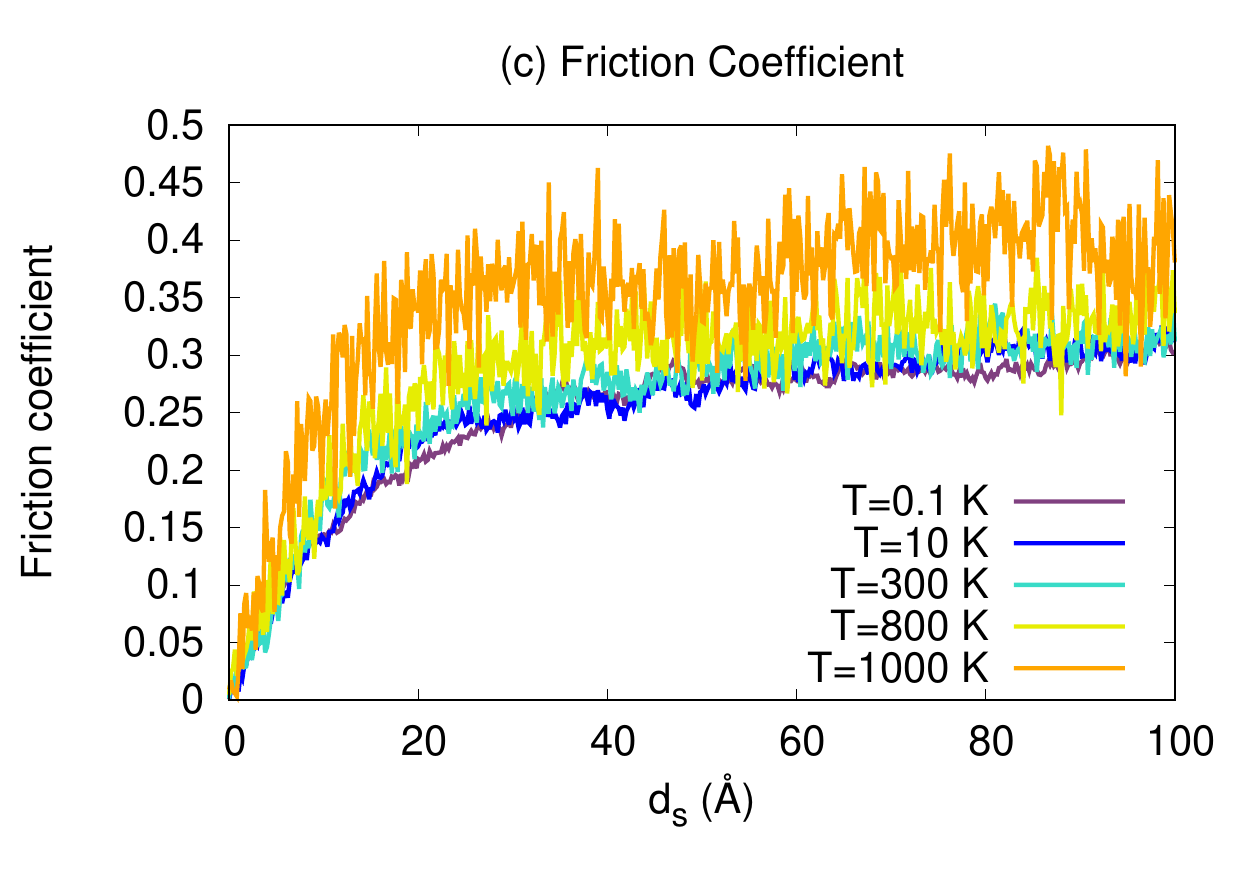}
    \includegraphics[width=0.45\textwidth]{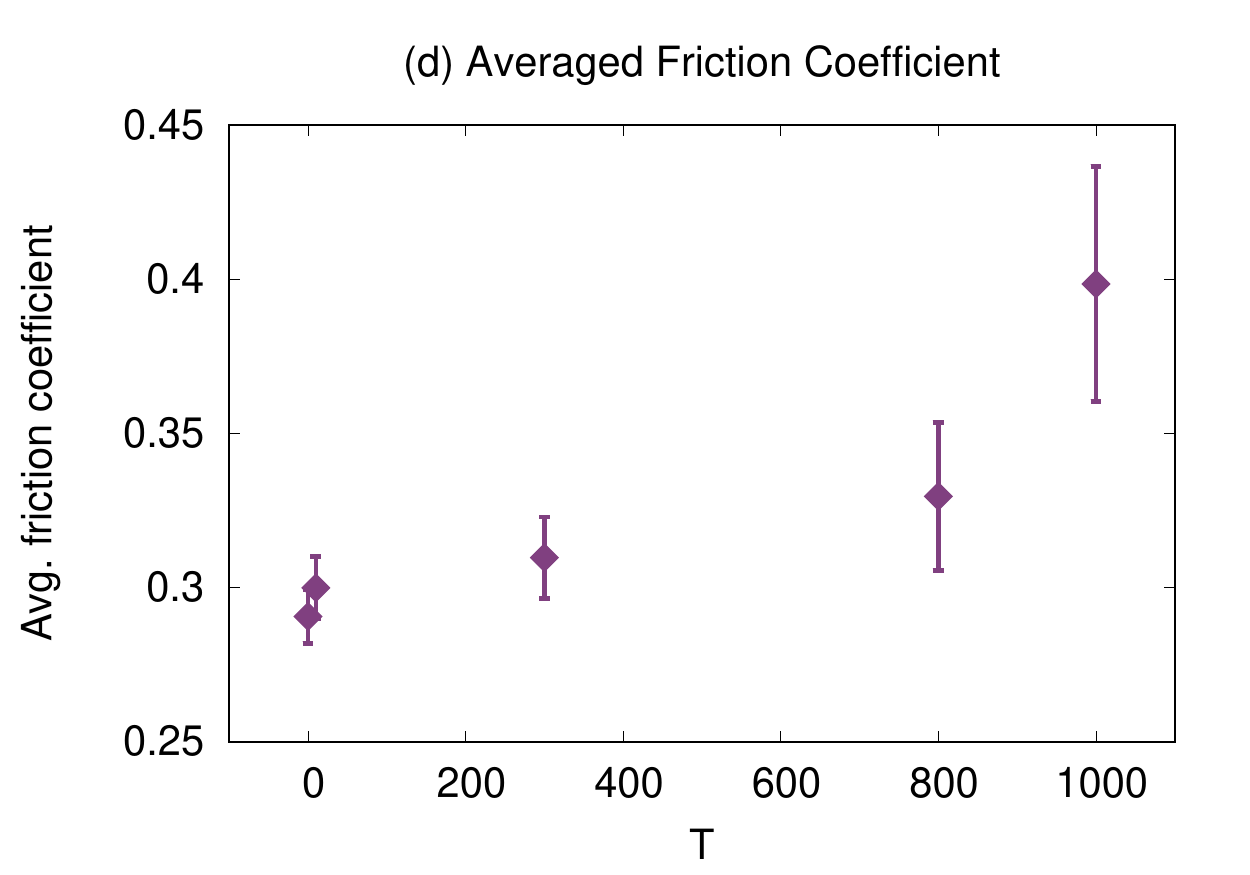}
  \end{center}
  \caption{(a) Normal force $F_n$ as a function of scratching distance, $d_s$,  for temperatures $T= 0.1$~K, $10$~K, 300~K, 800~K and 1000~K. (b) Tangential force $F_t$ as a function of scratching distance for the temperatures presented in (a). (c) Friction coefficient $\mu=F_t/F_n$ as a function of scratching distance for the temperatures presented in (a). (d) Friction coefficient as a function of temperature calculated from figure (b) for values of scratching distance $>$ 80~\AA. 
  }
  \label{fig:diff_t}
\end{figure*}

Figs.~\ref{fig:diff_t}a and b show these forces as a function of the scratching distance $d_s$ for various temperatures. The normal force decreases during the first $\sim$ 30 \AA\ and then becomes stationary. The first 30 \AA\ are the onset regime, where the system changes from the indentation to the scratch mode; while during indentation the entire submersed part of the tip has contact to the substrate and exerts a force on it, during scratch only the front part of the tip loads the substrate, while the rear partly loses contact; concomitantly the transmitted normal force decreases. This behavior is typical of scratching simulations and has been found also in crystalline materials~\cite{KCR00a,ZSHY12,GBKU15}. Also the transverse force shows such an onset behavior, Fig.~\ref{fig:diff_t}b; after 30 \AA, however, the force does not saturate, but shows a slight increase. This increase is caused by the frontal pileup forming in front of the scratch tip which increases in the course of scratching, see \qsect{s_pile} below, and hence leads to increasing transverse forces.

The (instantaneous) friction coefficient is readily
defined from these forces as
\beql{e_f}  \mu = \frac {F_t}{F_n}. \eeql 
As Fig.~\ref{fig:diff_t}(c) demonstrates, after the onset regime, it slightly increases with scratching distance, $d_s$. As $F_n$ is quite stationary, this increase is caused by the steadily increasing transverse force and hence by the forming pileup. As with the behavior of the normal force the evolution of the transverse force and of the instantaneous friction coefficient are generic and not restricted to glassy material.

All forces show a distinct temperature dependence in that they decrease with increasing temperature whereas the friction coefficient increases with increasing temperature. 
Fig.~\ref{fig:diff_t}(d) focuses on this behavior by plotting the friction coefficient averaged over the last 2 nm of scratching. 
If $F_n$ and $F_t$ displayed the same temperature dependence, $\mu$ would be insensitive to $T$. The fact that we observe a slight increase of $\mu$ with temperature hence means that the transverse force does not decrease as strongly with temperature as the normal force. Note that  the increase becomes particularly pronounced when $T$ rises above the glass temperature which is due to the activation of common thermal relaxation mechanisms as observed, for instance, in mechanical loss spectroscopy~\cite{KM17}.

In order to calculate the contact pressure and hence the hardness, we first need to determine the contact areas of the tip in normal and in transverse direction. Since we know the positions of all atoms with respect to the scratch tip, these areas can be determined atomistically from the atoms that are in contact with the indenter; for details see Appendix \ref{appendix2}.

\begin{figure*}
  \begin{center}
    \includegraphics[width=0.45\textwidth]{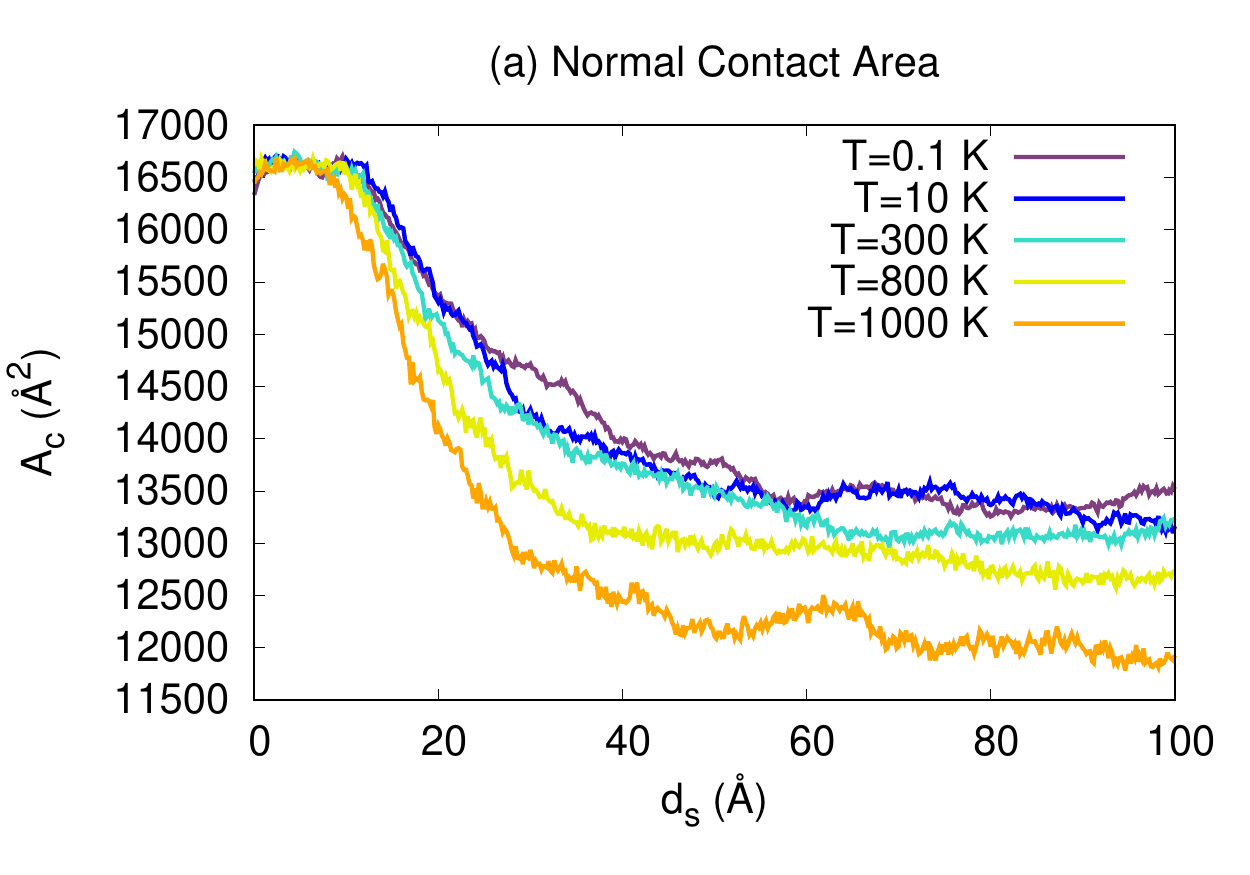}
    \includegraphics[width=0.45\textwidth]{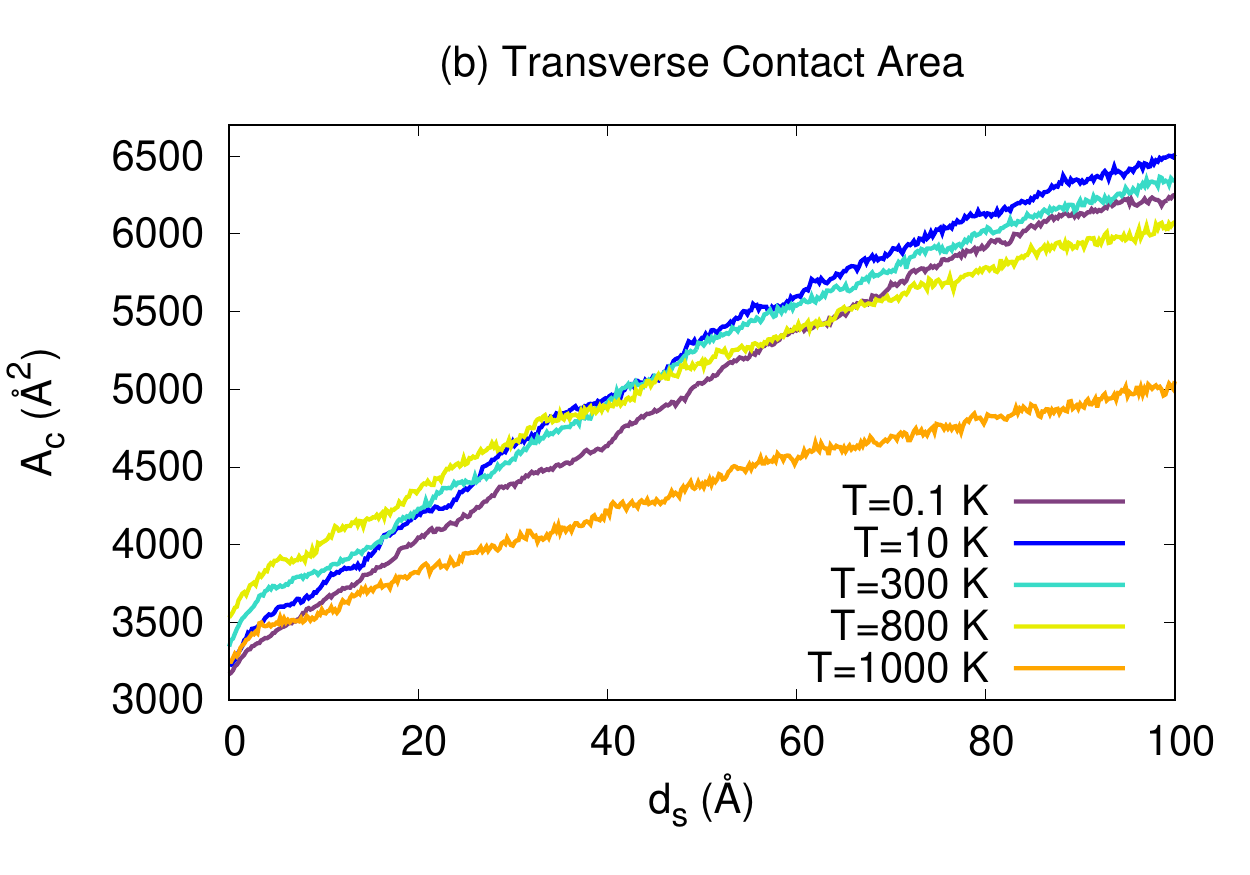}
  \end{center}
  \caption{Evolution of the (a) normal and (b) transverse contact area $A_c$  
with scratching distance, $d_s$, for various temperatures $T$.
  }
  \label{fig:Area}
\end{figure*}

The contact area normal to the surface, Fig.~\ref{fig:Area}(a),  decreases during the onset regime  by around 30 \%; as mentioned above this is caused by the tip losing contact with the substrate in its rear part. Macroscopic theories of scratching actually assume the normal contact area to decrease by 50 \%, such that only the front part of the moving spherical tip remains in contact with the surface~\cite{BT66}. For nano-sized tips this is different, since the elastic rebound of the groove floor behind the tip keeps part of the rear surface in contact with the tip.  

The transverse contact area,  Fig.~\ref{fig:Area}(b), shows a steady increase after the start of the scratching which is caused by the pileup increasing in size and contacting the tip surface.

The normal contact area shows a distinct temperature dependence; its values are systematically lower at higher temperatures.  During indentation, of course, the contact area is well defined and independent of temperature. Interestingly, it decreases only after a scratching distance of  $\approx$~10~\AA. This marks the point when the rear part of the indenter starts losing contact with the sample due to the  starting groove formation. 
While up to 300 K, the temperature sensitivity is negligibly, it becomes more pronounced for higher temperatures, and in particular  above the glass temperature. This is consistent with the rebound shown in Fig. \ref{fig:pileup}, a higher and immediate rebound would cause a higher contact area, which is the case for low temperatures. 

The transverse contact area shows almost no temperature dependence up to 800 K and only decreases for the highest temperature 1000 K. This is in line with the reduction of the pileup height at this temperature, see \qsect{s_pile} and Fig.~\ref{fig:pileup}.

The contact pressure can be determined from the force and contact area as 
\begin{equation} p_c = \frac {F}{A_c} . \label{Hardness} \end{equation}
We distinguish between a normal and a transverse (or scratching) contact pressure. Averaged values are denoted as the hardness.%The average over the contact pressure is denoted as the material hardness; in our case we average over the last 2 nm of the scratch process.
\begin{figure*}
  \begin{center}
      \includegraphics[width=0.45\textwidth]{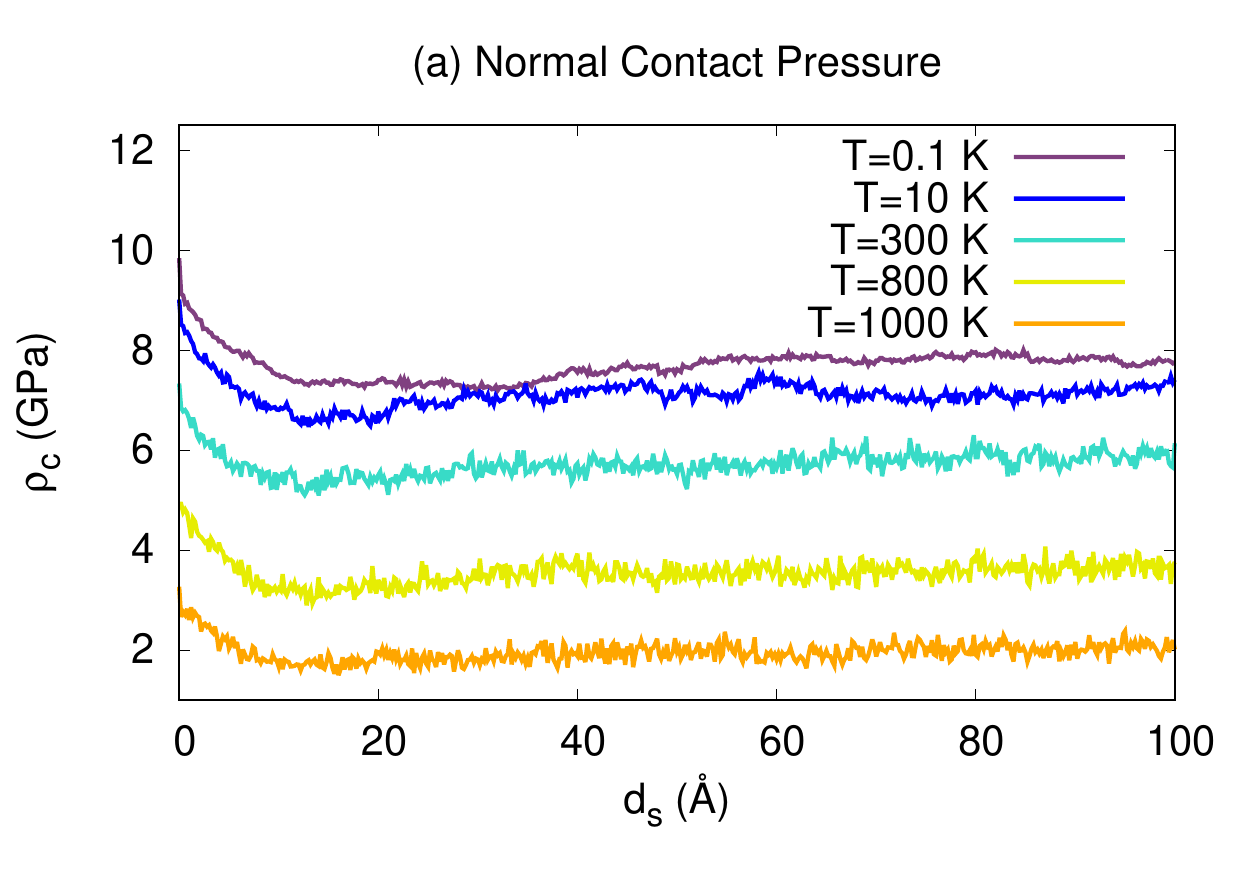}
    \includegraphics[width=0.45\textwidth]{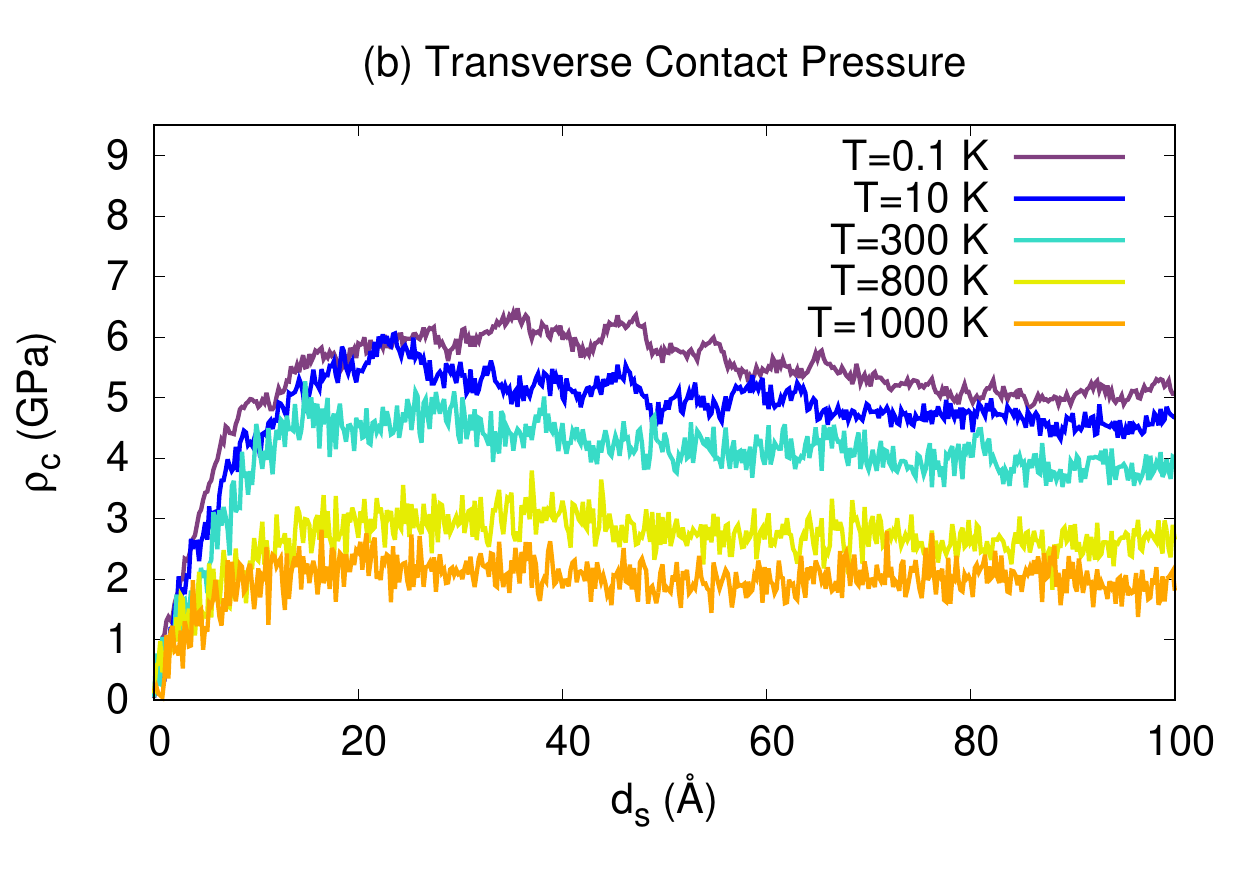} \\
    \includegraphics[width=0.45\textwidth]{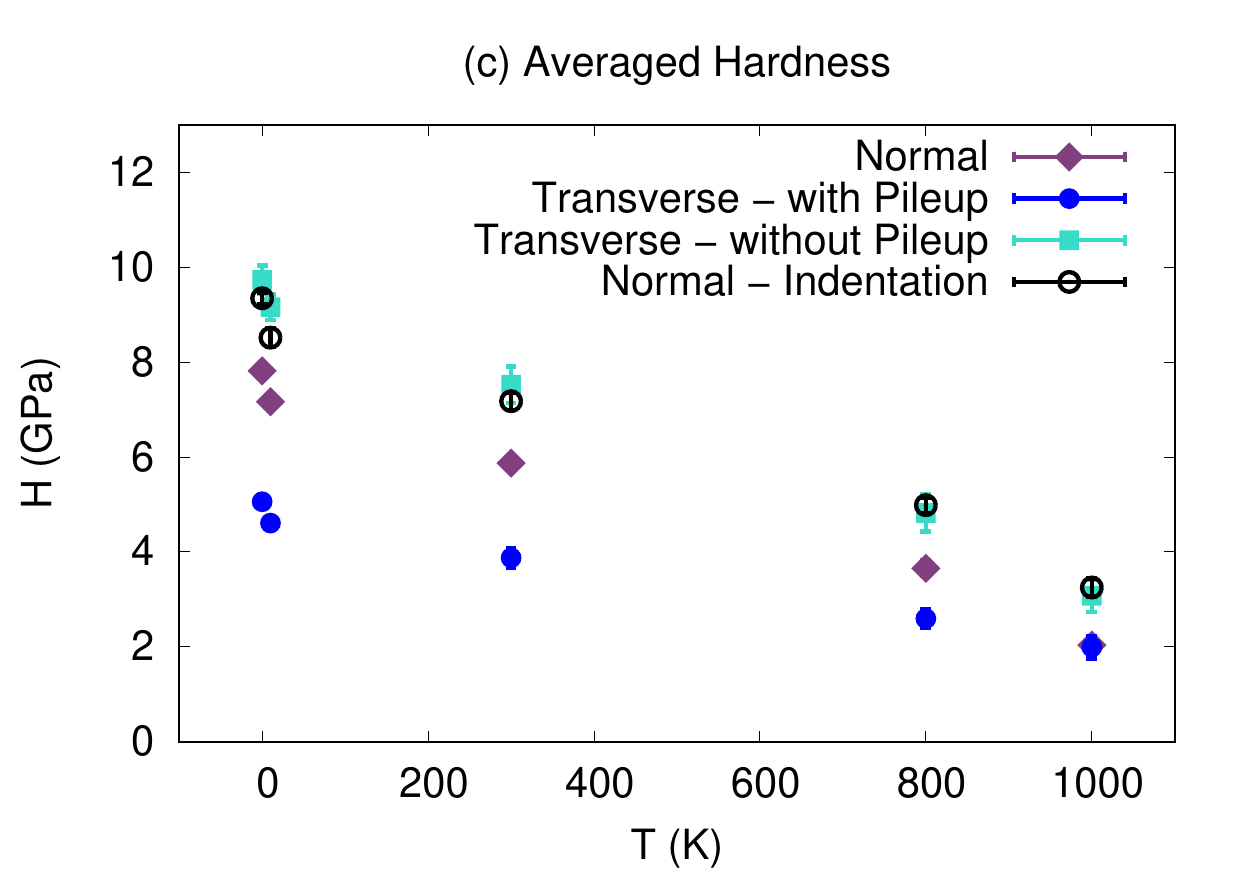}
   \end{center}
  \caption{ Evolution of the (a) normal and (b) transverse contact pressure $p_c$  
with scratching distance, $d_s$, for various temperatures $T$.
  (c) Normal and transverse scratching hardness as well as indentation hardness as a function of temperature. The indentation hardness  was averaged between indentation depths of 20 \AA to 30 \AA. The scratching hardnesses are calculated  as averages over the contact pressure for  scratching distances $>$ 80~\AA. For the transverse hardness, two values are provided, depending on whether the pileup above the surface was included in the are determination or not; see text.
  }
  \label{fig:Hardness}
\end{figure*}

Figs.~\ref{fig:Hardness}(a) and (b) show the evolution of the contact pressure with scratching distance. The normal contact pressure shows almost no variation; during the onset regime it appears to decrease slightly, showing that the normal pressure during scratching is somewhat lower than the normal pressure during indentation, see Fig.~\ref{fig:Hardness}(c). We attribute this feature to the more complex loading pattern during scratch, where the load is concentrated on the front part of the tip while the material under the rear part starts unloading. However, this effect is minor, changing the normal contact pressure by $\sim$ 1.5 GPa.

The transverse contact pressure, Fig.~\ref{fig:Hardness}(b), can only be discussed after the onset stage. We see that it attains quite stable values during scratching, even though the pileup increases steadily during scratch. This demonstrates that the transverse force increases in good proportionality with the transverse area. Note that the fluctuations in the transverse contact pressure  are considerably larger than for the normal contact pressure, in particular for low temperatures. 
The fluctuations come from the intermittent behavior of the force shown in Fig.~\ref{fig:diff_t}(b); they are particularly strong at low temperatures. This is already known from the stick-slip mechanism perspective~\cite{HWR*17,HWR*17a}, which implies that the load drops, or strain burst, are more pronounced at low temperatures~\cite{KPA*14}.

Both normal and transverse contact pressures show a strong dependence on temperature. Again we calculate the data  averaged over the last 2 nm of scratching, denoted as the hardness, and display it in Fig.~\ref{fig:Hardness}(c). The strong decrease with temperature seen is caused by a weakening of the glass at higher temperatures~\cite{Wan12a}.

The normal hardness is larger than the transverse hardness.
This is caused by the fact that the resistance of the material to the scratching tip has two sources: (i) the material below the original surface plane and (ii) the pileup. While the resistance of the material (i) will lead to similar values as the normal hardness, the pileup (ii) will be softer, since it only a thin bulge on the surface with a width of around 30 \AA,  see  Fig.~\ref{fig:Profile_view}. The reduced resistance of the pileup material thus causes the smaller values of the transverse hardness as compared to the normal hardness.

Note that in the calculation of the transverse hardness all substrate material contacting the tip  --~in particular the pileup~--  is included in the calculation of the transverse area. In experiment, often only the contact with the submersed part of the tip is included, since it can be easily determined from the groove shape. We add such a hardness determination in Fig.~\ref{fig:Hardness}(c); here the forces determined from  MD were taken unchanged while the area was calculated without including the pileup (not shown here). Indeed these data show strongly increased hardness values that are even larger than the normal hardness. 
Interestingly, the transverse hardness without taking the pileup into account is quite close to the normal hardness values calculated during indentation. This can be taken as an indication that this determination of transverse hardness -- which relies on the plastic resistance of the sub-surface material to the scratching tip --  
is not unreasonable.
We note that a similar conclusion could be drawn for the scratching of crystalline Fe in various surface orientations and scratch directions~\cite{GBKU15}.

\subsection{Active zones}  \label{s_STZ}

\begin{figure*}
  \begin{center}
    \includegraphics[width=0.95\textwidth]{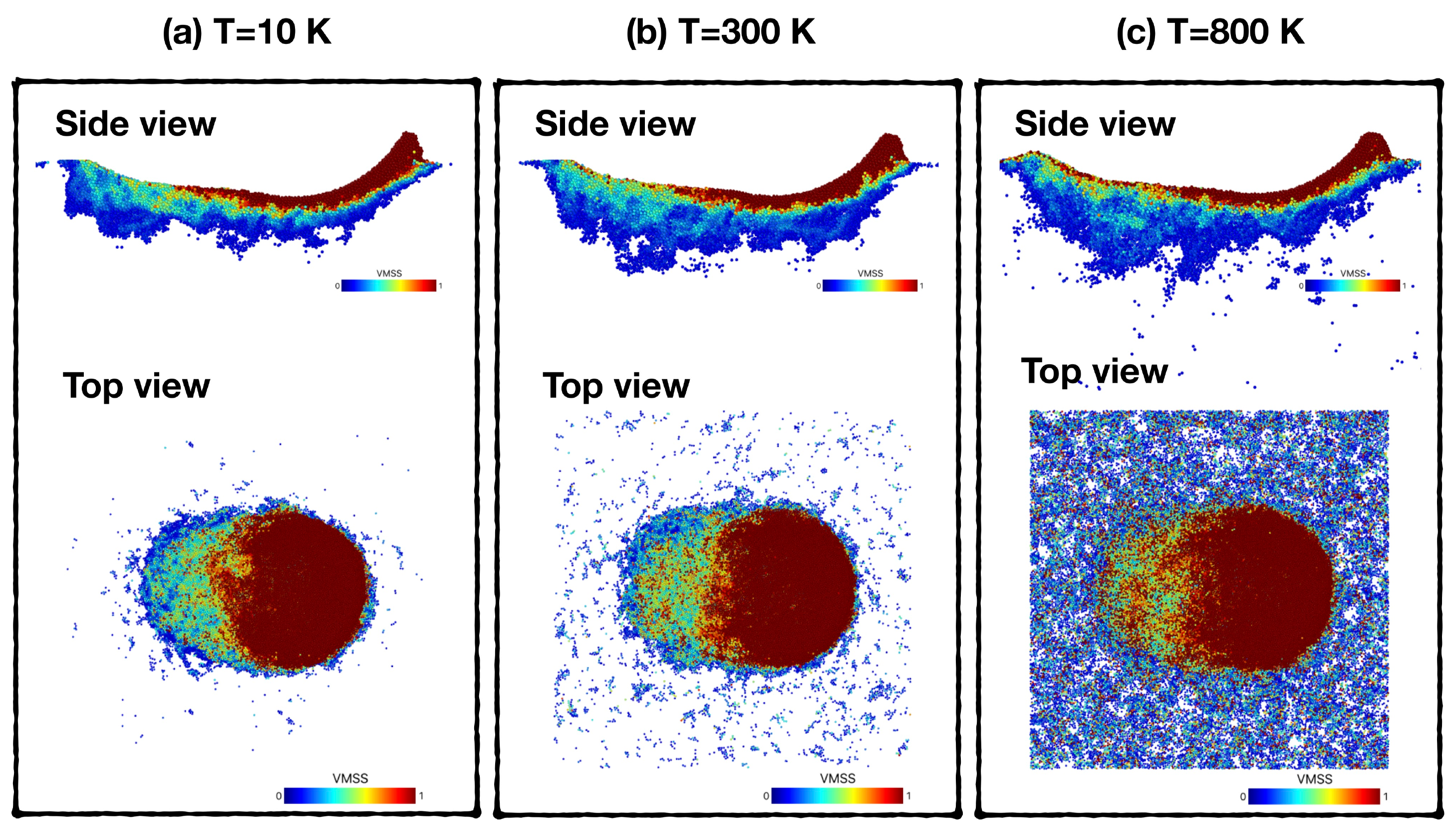}
  \end{center}
  \caption{Von-Mises shear strain (VMSS) snapshots of a thin slab ($\sim20$~\AA) and a top view of the sample at temperatures (a) $T=10$~K, (b) $T=300$~K and (c) $T=800$~K. These snapshots correspond to the maximum scratching distance 100~\AA\ with virtual indenter still in the sample and shows only atoms with VMSS$>0.1$.
  }
  \label{fig:VMSS_diff_t}
\end{figure*}

Now, we want to determine the damage induced inside the sample by the scratching procedure performed in the surface.
Fig.~\ref{fig:VMSS_diff_t} shows the deformation due to the initial indentation and the succeeding scratching
for temperatures T=10 K, 300 K and 800 K for a scratching depth of 30~\AA. Here, we analyze the von-Mises shear strain (VMSS), which is the conventional way to analyze glass plasticity in atomistic simulations~\cite{SOL07}. We deleted the atoms with VMSS less than 0.1 to better observe the  zone where plastic activity occurs. For each temperature, the figure shows a side view of a thin slab (thickness of around 20~\AA) and a top view snapshot. These figures are taken after completing the entire scratching distance of 100~\AA\ when the indenter has not yet been withdrawn. The set of particles shown in Fig.~\ref{fig:VMSS_diff_t} we name the active zone. We want to emphasize that this set of particles shown in this figure, also contains particles that can, in principle, return elastically or visco-elastically to their original configuration with the release of the indenter pressure~\cite{AKAU19}. However, since we are not quantifying strain irreversibility, we might sometimes refer to the active zone as the plastic zone, without ambiguity. 
The active zone at this depth, as shown in this figure, mostly consists of STZs. These STZs consist of such atoms that move cooperatively and thus form regions with high shear strain. Only at low temperatures larger STZs start forming in string-like arrangements pointing toward the beginning of shear band formation.
Also, the active zone increases with temperature, suggesting more localized plastic events for low temperatures. This effect is well known and has been reported in nanoindentation studies~\cite{SLN04,ZZZW17,AKAU19} and other loading experiments~\cite{GCM13,HSF16,RNTR18}.  From the top view snapshots, a very noticeable activity can be observed at the surface of the sample with increasing temperature, with no shear-band formation at the surface at any of the studied temperatures.

\begin{figure*}
  \begin{center}
  \subfigure[]{\includegraphics[width=0.45\textwidth]{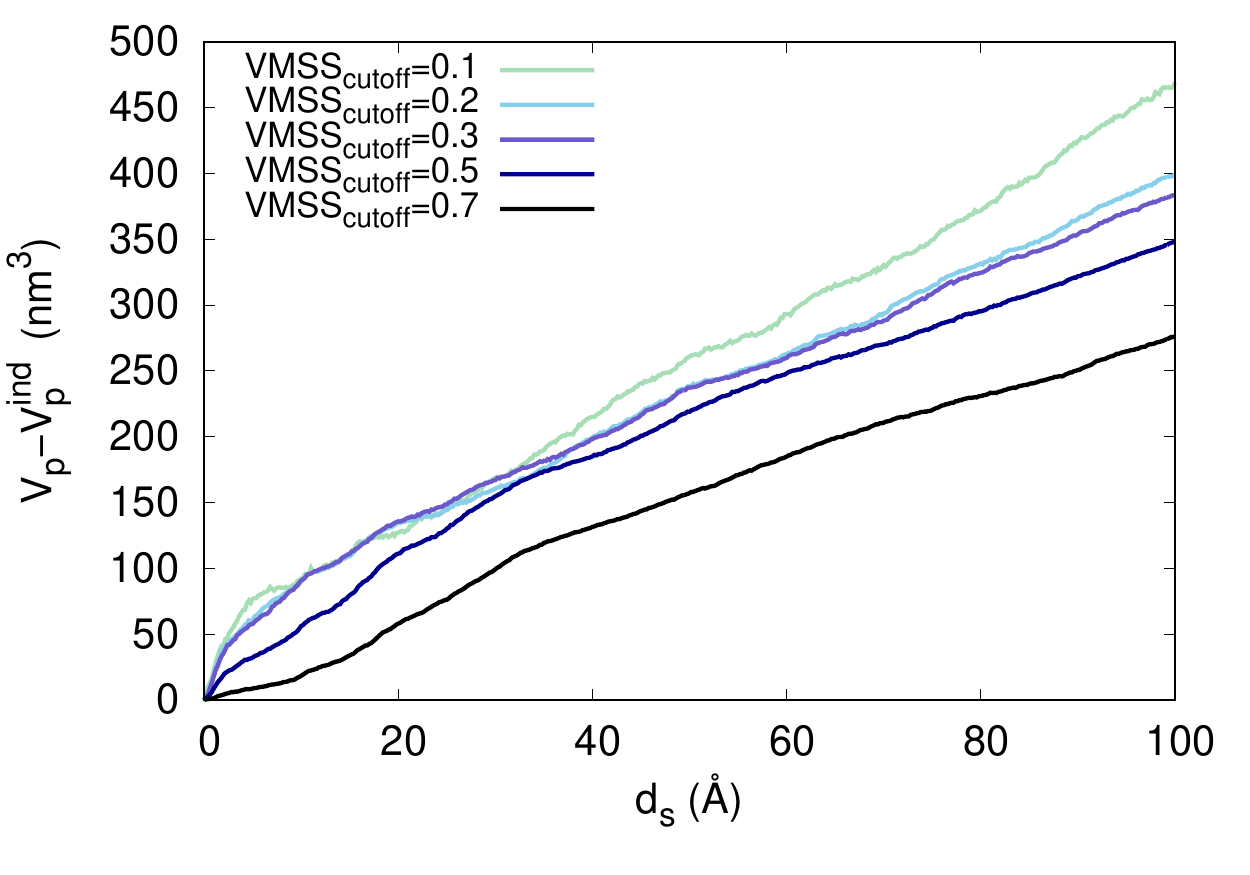}}
      \subfigure[]{\includegraphics[width=0.45\textwidth]{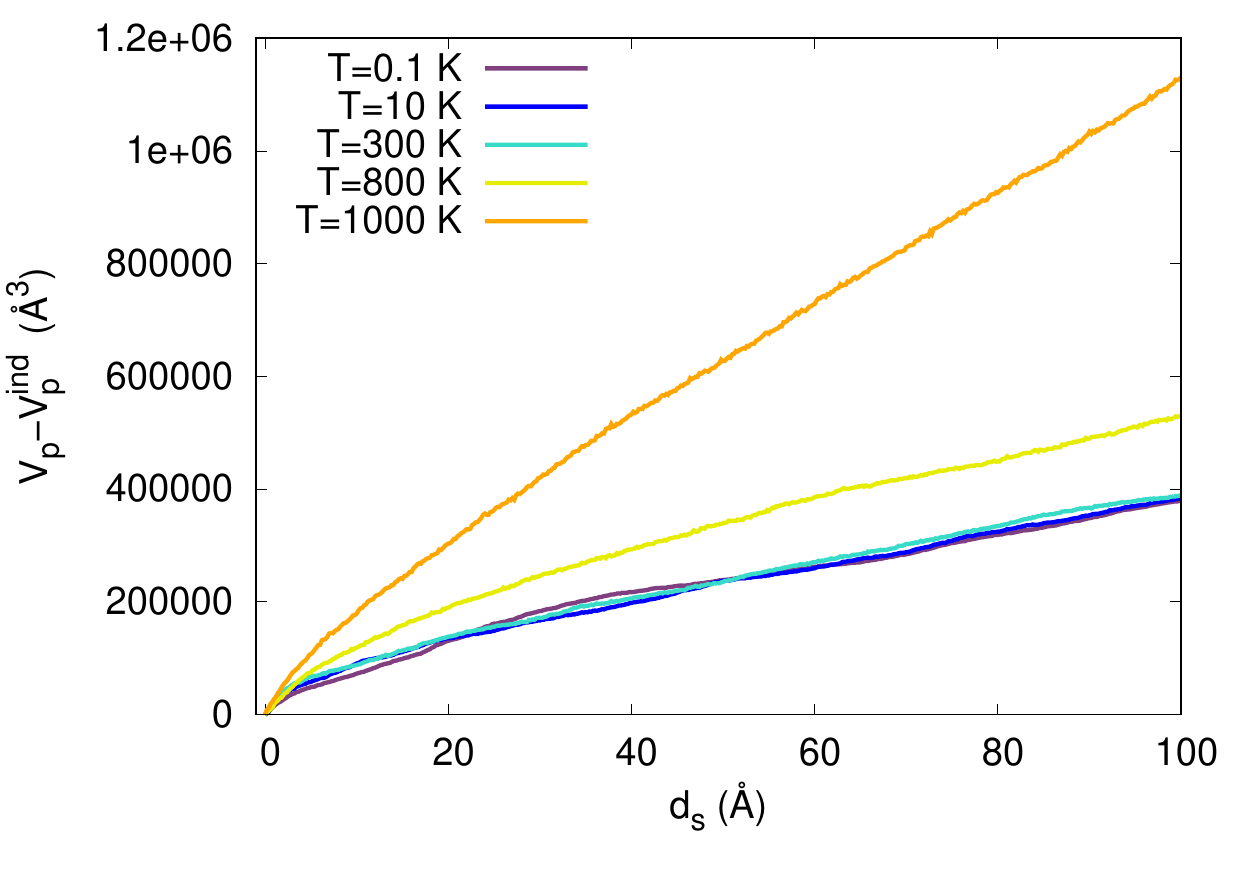}}\\
      \subfigure[]{\includegraphics[width=0.45\textwidth]{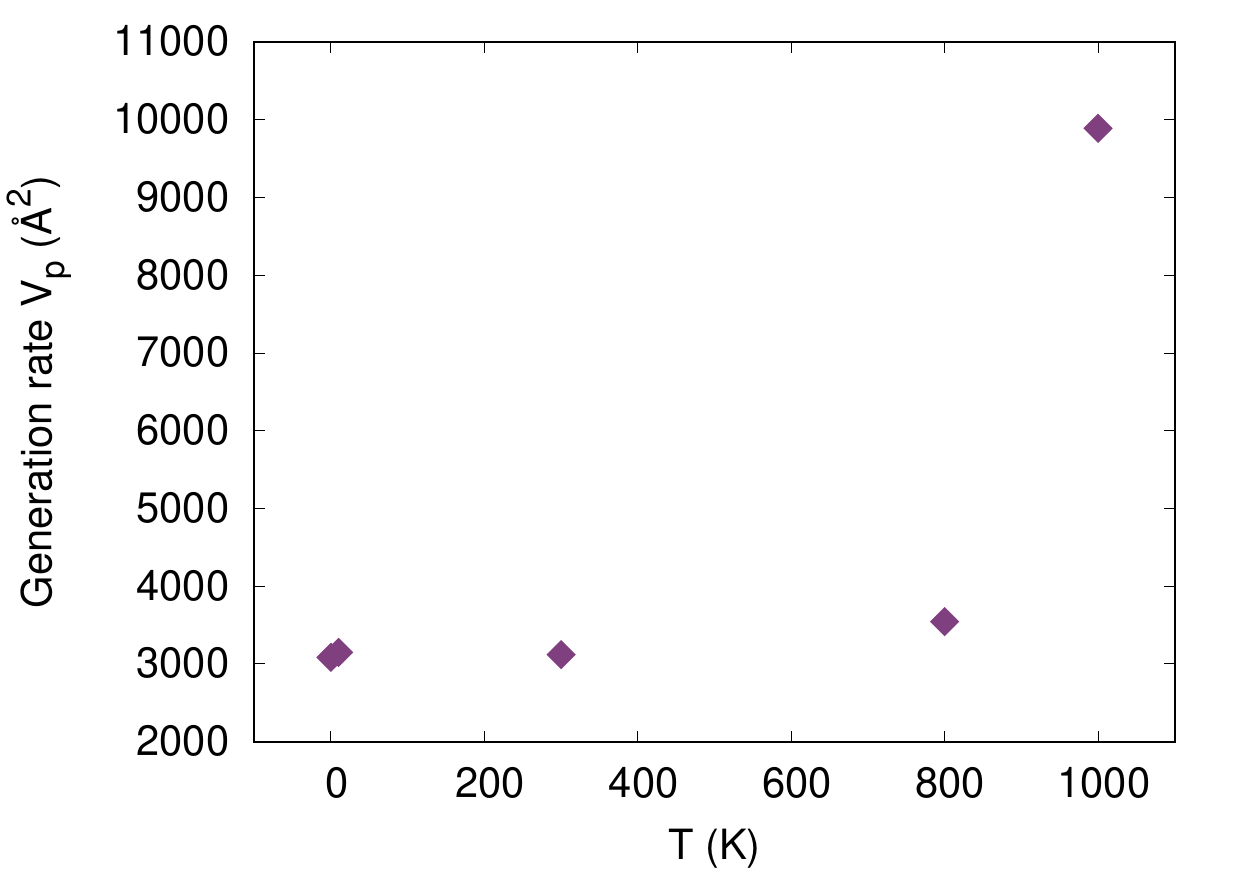}} 
  \end{center}
  \caption{(a) Volume of the active zone, $V_p$, as a function of scratching distance at $T=10$ K  for different values of VMSS$_{\text{cutoff}}$. The  volume created during indentation, $V_p^{\rm ind}$, has been subtracted. (b) $V_p$ as a function of scratching distance for different temperatures. Note the difference in ordinate scale in panels (a) and (b). (c) Generation rate of active atoms, ${\rm d}V_p/{\rm d}d_s$, as a function of temperature. }
  \label{fig:Plasticity}
\end{figure*}

In order to quantify the plasticity produced during scratch, we monitor the atoms that underwent large von-Mises strains~\cite{SOL07},  as seen in Fig.~\ref{fig:VMSS_diff_t}.  Shear transformation zones (STZs) consist of such atoms that move cooperatively  and thus form regions with high shear strain.   
We denote by $N_p$ the number of atoms with large value of the VMSS.  For this analysis we exclude the atoms belonging to the pileups and focus on the material below the original surface. We convert $N_p$ to the volume of the active zone, $V_p = N_p\Omega$, by multiplying with the (weighted) atomic volume of our system, $\Omega= 16.1$~\AA$^3$. 
Since we are interested in the effect of scratching, the volume created during indentation, $V_p^{\rm ind}$,  has been subtracted.  Technically, we need to introduce a cut-off value of the von-Mises strain, VMSS$_{\text{cutoff}}$, such that  $V_p$ quantifies the number of atoms with von-Mises strain above VMSS$_{\text{cutoff}}$. It has been argued that a reasonable value is around 0.2~\cite{ZZC*16,ZCW*17}.  We monitor the sensitivity of $V_p$ on VMSS$_{\text{cutoff}}$ in Fig.~\ref{fig:Plasticity}(a). It can be observed that the growing behavior is very similar for all choices of VMSS$_{\text{cutoff}}$. We therefore select an intermediate value VMSS$_{\text{cutoff}}=0.3$ to perform a temperature--dependence analysis.

Fig.~\ref{fig:Plasticity}(b) shows how the volume of  the plastic zone increases as a function of distance for different temperatures. For the high temperature there is a more rapid growth of the active zone as a function of scratching distance than for lower temperatures. This behavior is due to the enhanced relaxation due to homogenous flow.%This behavior at high temperature is due to the enhanced relaxation which leads to a more homogenous deformation of the sample, which means that each volume element undergoes similar strain.

The volume generation rate of the atoms  of the active zone, obtained from the slope of Fig.~\ref{fig:Plasticity}(b), is shown in Fig.~\ref{fig:Plasticity}(c) as a function of temperature. The numbers show a slight trend of increase with $T$, but it is pronounced only above the glass transition temperature.

\begin{figure*}
  \begin{center}
  \subfigure[]{\includegraphics[width=0.45\textwidth]{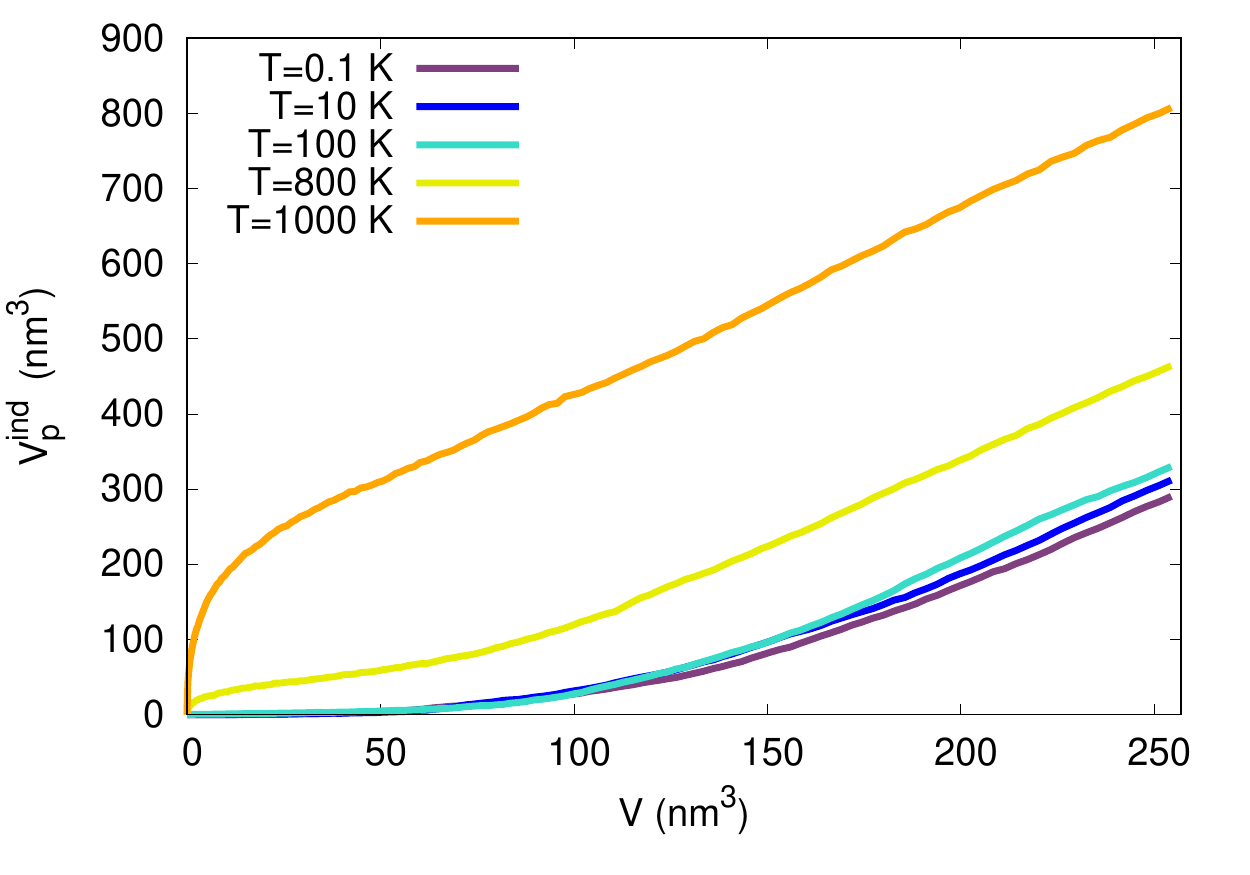}}
      \subfigure[]{\includegraphics[width=0.45\textwidth]{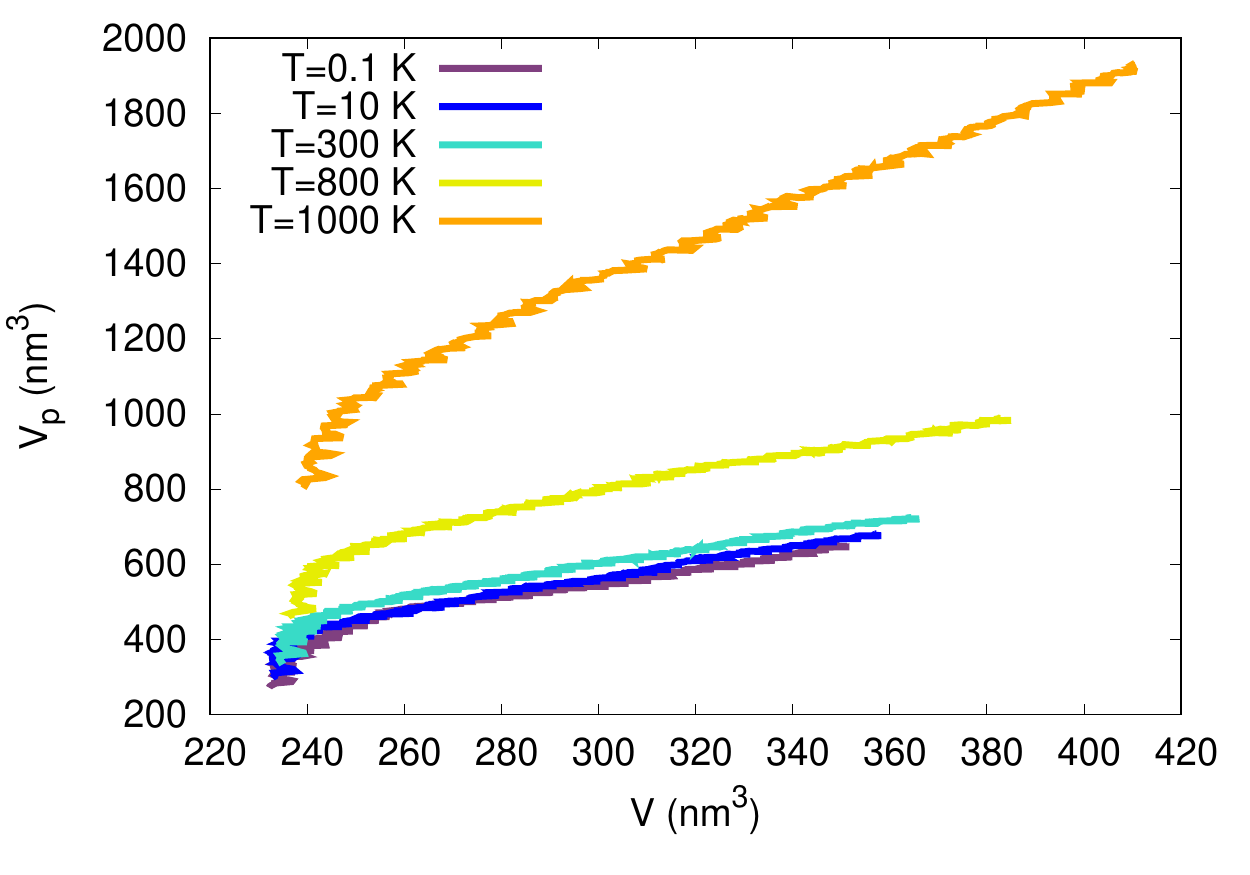}}\\
      \subfigure[]{\includegraphics[width=0.45\textwidth]{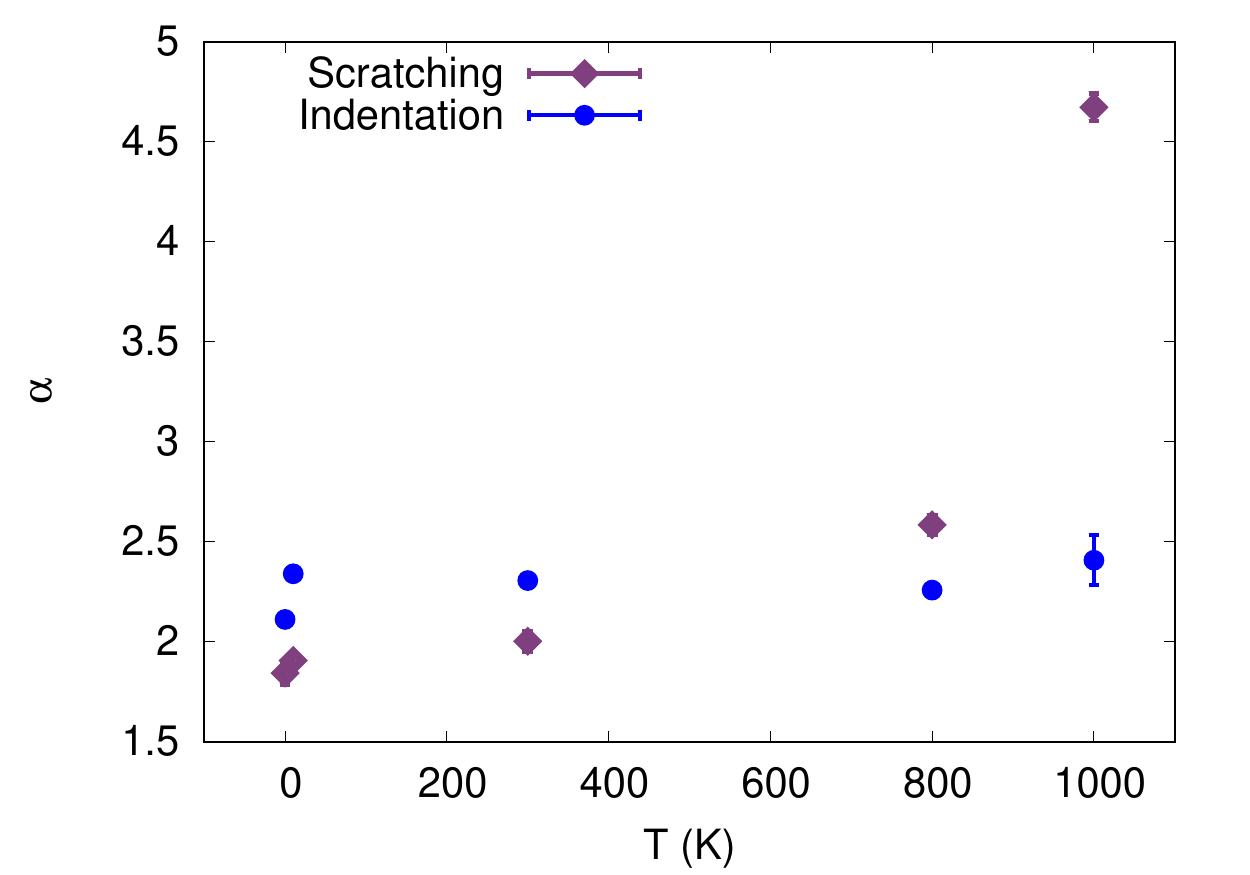}} 
  \end{center}
  \caption{
Volume of the active zone, $V_p$, generated after (a) indenting and (b) scratching a volume $V$. Note the difference in ordinate scale in panels (a) and (b). (c) Plastic efficiency, $\alpha$, \qeq{e_eff},  as a function of temperature. }
  \label{f_NV}
\end{figure*}

Finally, we want to compare the active volume produced during scratch with the one produced during the indentation process. In order to put the comparison on the same footing, we  plot the active volume, $V_p$, versus the volume excavated $V$ during indent, \qfig{f_NV}a, or scratch, \qfig{f_NV}b. The volume of the indentation pit was calculated geometrically as a spherical cap. Since the volume of the scratching groove can in principle change, we re-measured the volume at every step using an integration method, similar  to the contact area calculation, see Appendix \ref{appendix2}.
 In both cases, an approximate proportionality of $V_p$ with  $V$ is observed. This allows us to define a plastic efficiency of the indentation and grooving process as 
\beql{e_eff} \alpha = \frac {V_p} {V} . \eeql
 \qfig{f_NV}c demonstrates that the efficiency is in both cases of the order 2; this means that as the consequence of each atom that is removed from the substrate by machining, the active volume  increases by 2 atomic volumes.  
 This information is reassuring since it means that from the point of view of the total plastic volume generated, the kinematically simpler and computationally less costly process of nanoindentation can be used for future studies in this field, rather than the more complex scratching process. In detail the plastic efficiency is slightly 
higher for indentation than for scratching for lower temperatures. However, at high temperatures the scratching efficiency overtakes the one for indentation meaning that the thermal processes initiated during indentation continue and increase at high temperatures. The effect at low temperatures
 is caused by the fact that during indent the material is constantly under stress, while during scratch only the front part of the groove is stressed whereas  the rear part of the groove is starting to relax already, decreasing the active volume there, see \qfig{fig:VMSS_diff_t}. Finally we note that the plastic efficiency shows only little temperature dependence for $T \le 800$ K.  only when the glass temperature is approached, the plastic efficiency increases; the increase is  particularly strong for nanoscratching.

It can also be observed in \qfig{f_NV}b that the final volume is smaller at lower temperatures, meaning that there is more elastic recovery, which is consistent with Fig.~\ref{fig:pileup}. This recovery not only occurs at the groove floor, but also at the walls of the groove.

Overall, the results in Sec. \ref{s_pile},  \ref{s_F} and  \ref{s_STZ} show that the lower the hardness, the higher the friction coefficient and the generation rate of the plastic zones inside the sample. This in turn leads to a lower rebound, meaning lower elastic and visco-elastic response with increasing temperature. This is in accordance with experimental observations during dynamic mechanical excitations~\cite{HBS*08}, where it was found that the elastic response (real part of Young's modulus) decreases and the mechanical loss (imaginary part of Young's modulus) increases when approaching the glass transition temperature. This suggests that the increase in the friction coefficient is caused by enhanced activation of atomic rearrangements at higher temperatures.

\subsection{Scratching depth dependence} \label{s_depth}

\begin{figure*}
  \begin{center}
    \includegraphics[width=0.95\textwidth]{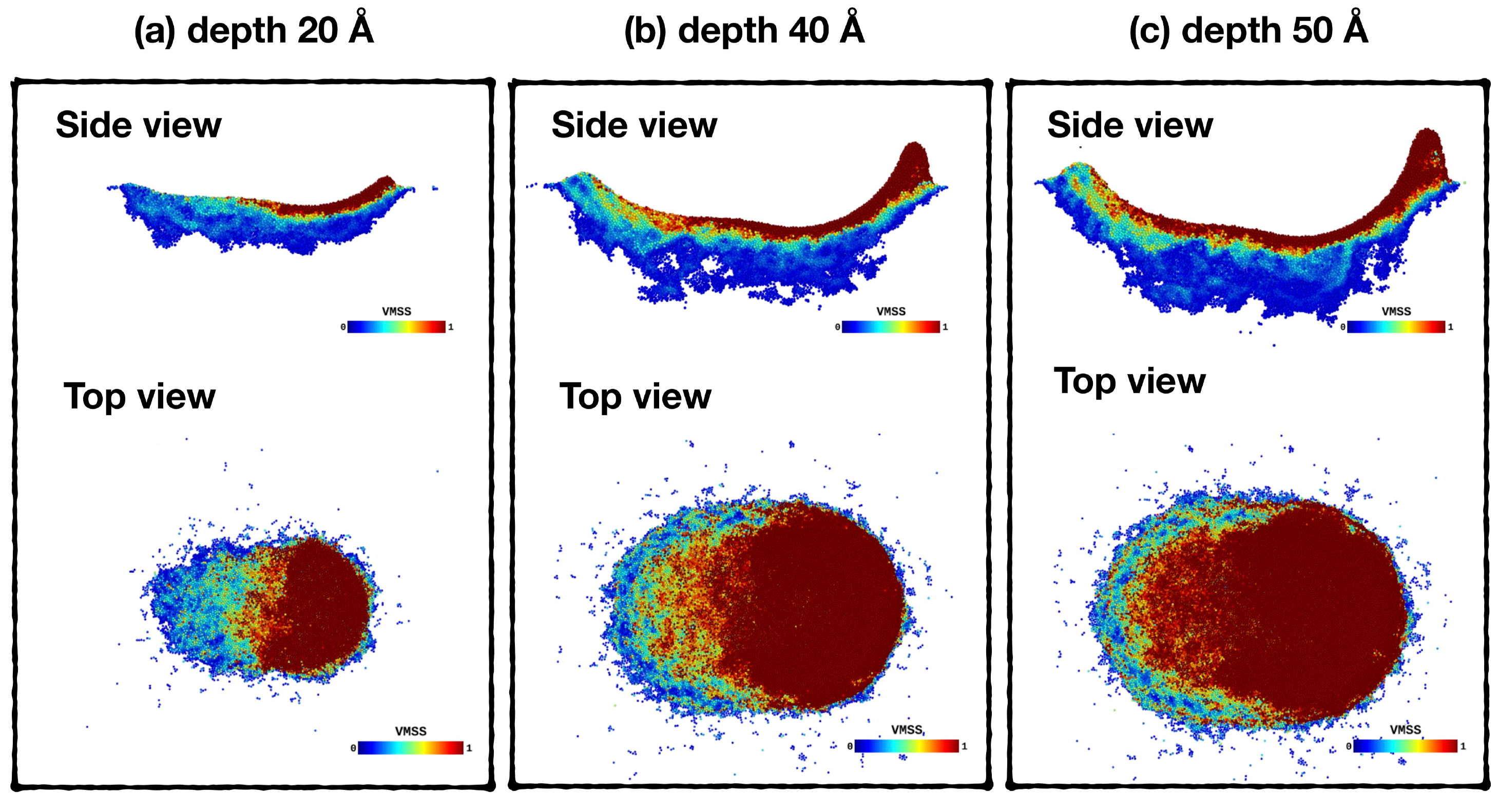}
  \end{center}
  \caption{Von-Mises shear strain (VMSS) snapshots of a thin slab ($\sim20$~\AA) and a top view of the sample at scratching depth (a) 20~\AA, (b) 40~\AA\ and (c) 50~\AA. These snapshots correspond to the maximum scratching distance 100~\AA\ with virtual indenter still in the sample.
  }
  \label{fig:VMSS_diff_depth}
\end{figure*}

In principle, a larger deformation could involve a different mechanism of deformation, i.e., a transition from STZs to shear band operation. Therefore, in this section, we are going to review the effects of shear band activity on the quantities measured in the previous section. In order to do this, we apply a larger deformation by increasing the depth of the initial indentation before scratching. Here, we keep a constant temperature $T=10$~K and scratch the sample at depths 20, 30, 40, 50 and 70~\AA. Fig.~\ref{fig:VMSS_diff_depth} shows the von-Mises shear strain (VMSS) for scratching depths 20, 40, and 50~\AA, including only atoms with VMSS $>0.1$. For each case, we show a side view of a thin slab (thickness of around 20~\AA) and a top view snapshot. One can observe that at scratching depth 50~\AA, there are shear bands already forming inside the workpiece and also in its surface. It can be seen that the shear bands in the surface propagate along the scratching direction and not perpendicular to it; this corresponds to experimental observations where shear bands are also seen to grow perpendicular to the scratch direction~\cite{HN04,HCS*10,KCL*08}.  This is also the case for scratching depth 70~\AA\ (not shown here).   

\begin{figure*}
  \begin{center}
      \subfigure[]{\includegraphics[width=0.45\textwidth]{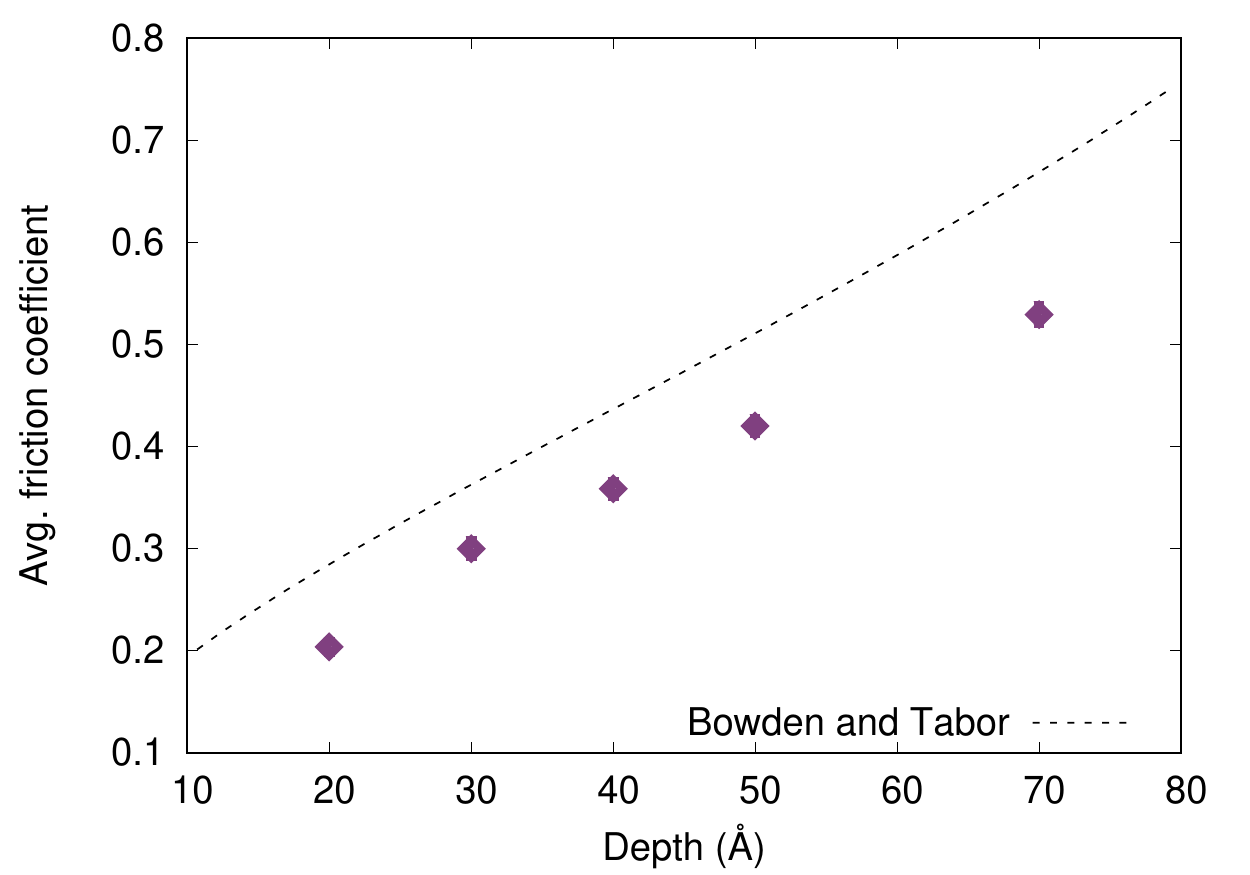}}
      \subfigure[]{\includegraphics[width=0.45\textwidth]{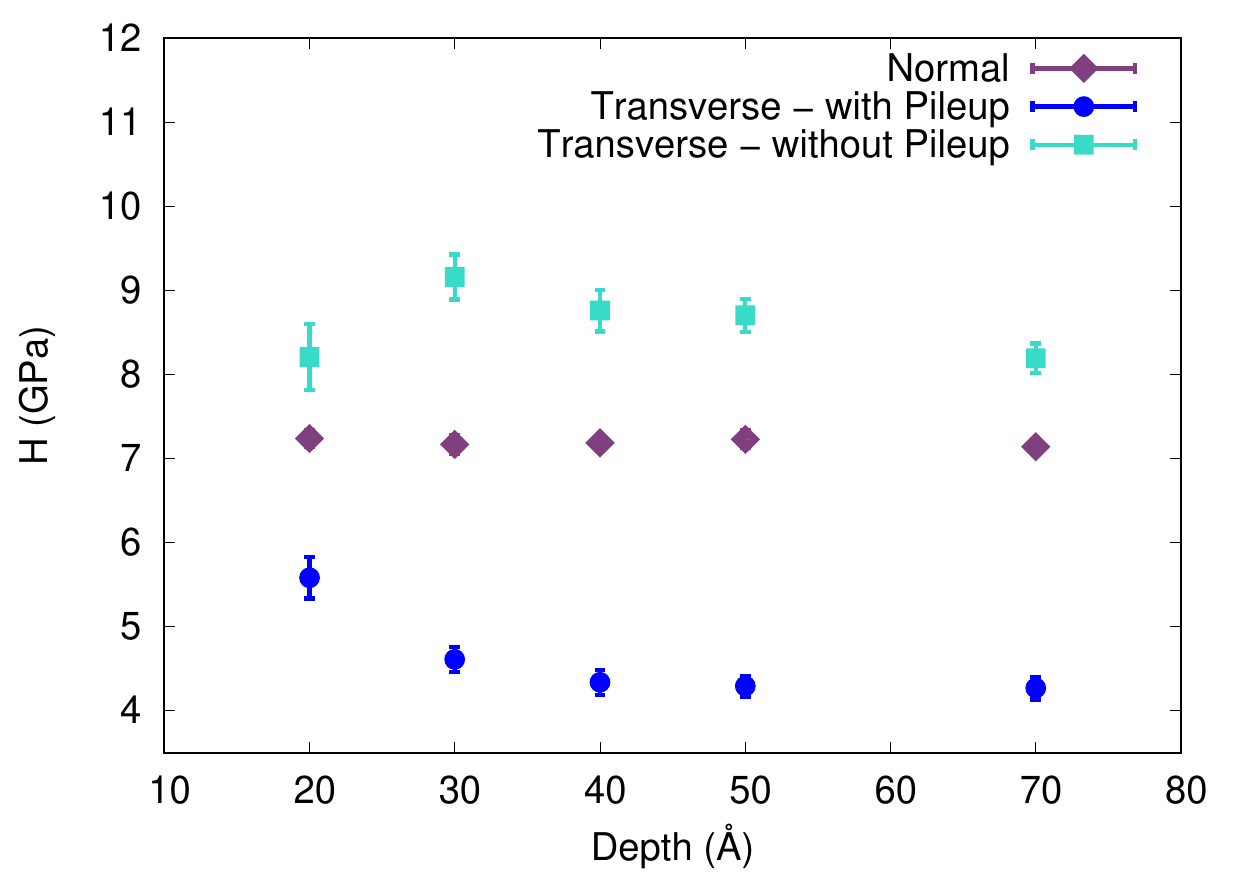}}\\
   \subfigure[]{\includegraphics[width=0.45\textwidth]{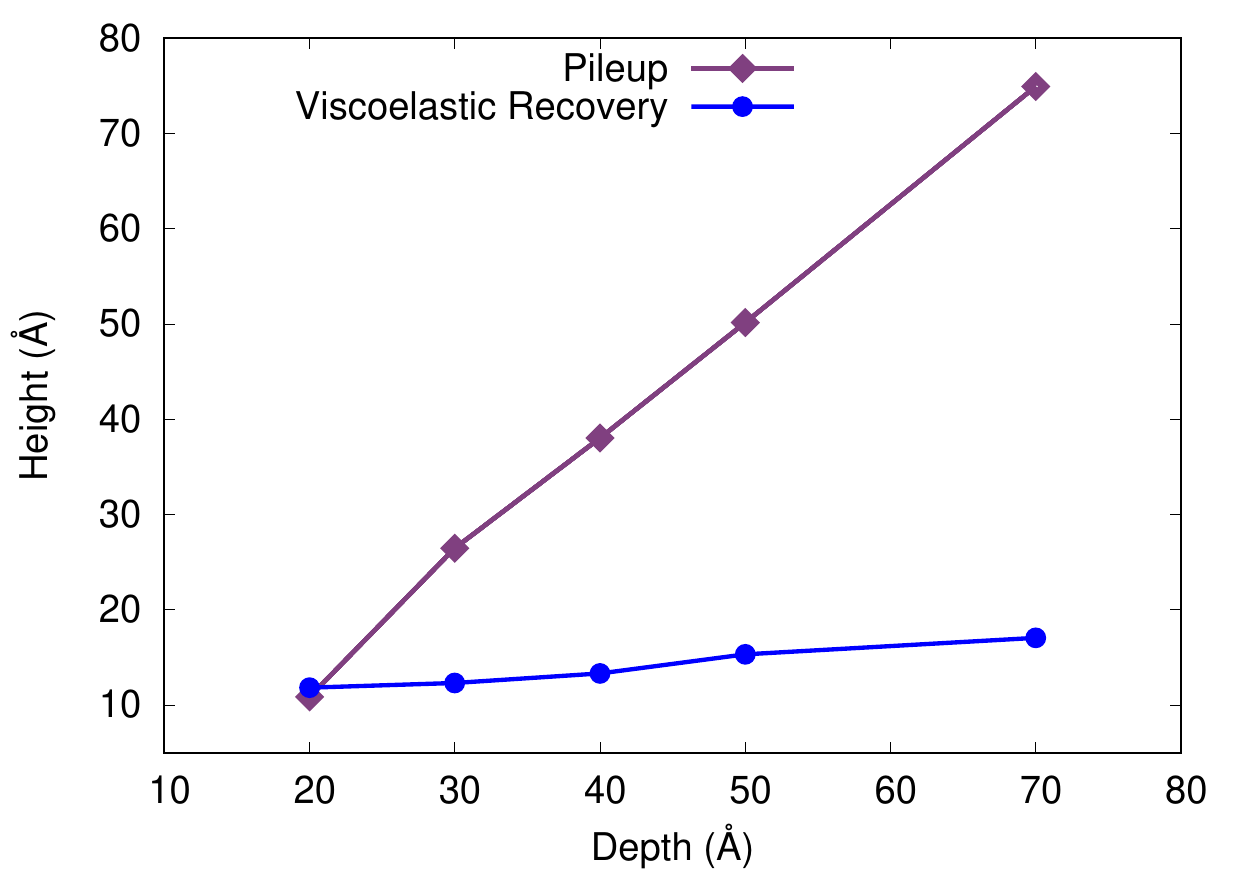}}
  \end{center}
  \caption{(a) Averaged friction coefficient  $\mu=F_t/F_n$ as a function of scratching depth (averaged over scratching distance $>$ 80~\AA). The line shows the geometrical prediction by Bowden and Tabor~\cite{BT66}, \qeq{e_BT}. (b) Normal and transverse hardness as a function of scratching depth. (c) Pileup and visco-elastic recovery height for different scratching depths. 
  }
  \label{fig:diff_depth}
\end{figure*}

Fig.~\ref{fig:diff_depth} summarizes the results obtained for various scratching depths. The coefficient of friction increases with scratching depth, see Fig.~\ref{fig:diff_depth}(a), in agreement with experiments~\cite{MZW*08}. The trend seems to be almost linear for depths $\ge$ 30~\AA. 

Bowden and Tabor~\cite{BT66} provided a simple geometrical prediction for the friction coefficient. Assuming that the normal and transverse hardness are identical, the friction coefficient is simply given by the ratio of normal and transverse areas of the submersed part of the tip. For a sphere this leads to a law
\beql{e_BT} \mu = \frac {2\theta - \sin (2\theta)}{\pi \sin^2 \theta}  \, , \eeql
where, $\theta$ is the semi-angle at the center of the spherical indenter subtended by the groove which is give as a function of the indenter radius and the scratch depth $d$ as
\beql{e_theta}  \cos  \theta = \frac{R-d}{R}. \eeql

As \qfig{fig:diff_depth}(a) shows, the geometrical model nicely explains the increase of the friction coefficient with scratch depth. Our values are, however, systematically smaller than the model; this is caused by the fact that the transverse hardness is smaller than the normal hardness: hence the area ratio overestimates the force ratio.

The normal hardness, shown in Fig.~\ref{fig:diff_depth}(b), does not depend on scratching depth. We note that the indentation hardness measured during  indentation to a depth of 30 \AA\ amounts to 8.52 GPa, see Fig. \ref{fig:Hardness}(c), very close to the normal scratching hardness.  The slightly smaller value during scratch can be associated with the more complex loading profile: while during indent the entire sample below the indenter is loaded, during scratch the front part is loaded while the rear part is unloaded. This more complex loading pattern results in a slightly smaller normal hardness during scratching.

However, the transverse hardness shows a pronounced depth dependence, see Fig.~\ref{fig:diff_depth}(b). This is connected to the growing frontal pileup, which provides less resistance to the moving tip than the sub-surface material. Indeed, as  Fig.~\ref{fig:diff_depth}(c) demonstrates, the pileup height increases more or less in proportion with the scratch depth. At scratch depths beyond around 4 nm, the transverse hardness saturates; evidently the intrinsic hardness of the pileup material is of the order of 4 GPa such that the lateral hardness cannot fall below this value. 
Again, as in the case of Fig.~\ref{fig:Hardness}(c), we have included in Fig.~\ref{fig:diff_depth}(b) the calculation of transverse hardness as done in experiments, i.e., excluding the pileup formation in the  determination of  the contact area. As discussed before in the text, using such a determination  of the contact area  overestimates the transverse hardness.

In Fig.~\ref{fig:diff_depth}(c) we also plot the height of the visco-elastic recovery. It also increase slightly, from around 11.8 \AA\ at a scratch depth of $d=2$ nm to 17.1 \AA\ for $d=7$ nm. This is plausible, since for a spherical tip, a larger scratch depth implies a larger tip contact area, and hence a larger volume that is loaded under compressive stress; hence unloading leads to a larger elastic and visco-elastic rebound.

The mechanical properties discussed  in this section show very little difference on the scratching depth, while the tribological properties show a  dependence proportional to depth. Although we observe shear bands at high scratching depth, our results suggest that the deformation is not high enough to observe changes in the macroscopical quantities.

\section{Summary}

In this work we used  the method of  molecular dynamics simulation to study the tribological properties of metallic glasses. 
This class of materials is known to exhibit a large elasticity; this feature has clear advantages for their tribological response, and in particular for the external damage of the material during abrasive wear, which is often analyzed in experiments. The total strain is distributed among a large elastic, visco-elastic and plastic strain, and so, in comparison to crystals, a given total strain can be more effectively accommodated at lower temperatures and partially recovered leading to a reduced external damage.

Our simulations show the creation of STZs during abrasive wear, similarly to other loading mechanisms such as nanoindentation. However in contrast to nanoindentation, the total volume of the rearranging regions exhibits a temperature dependence which leads to the surprising effect that metallic glasses are apparently more damage-tolerant during scratching than during nano-indentation at low temperatures. Well above room temperature, this effect is reversed.

We summarize the main findings of our atomistic study on plastic activity during nanoscratching of a CuZr glass as follows.

\begin{enumerate}

\item The groove shows a considerable amount of (visco-) elastic recovery of the order of 1--2 nm  for scratch depth between 2 and 7 nm. The pileup generated above the surface increases steadily with scratch direction reaching its summit immediately in front of the scratch tip.

\item After indentation, the system experiences an onset phase, after which scratching shows constant features (in normal direction). The normal force and contact area decrease during this onset phase while the rear part of the tip loses contact with the groove floor.  After  the transverse force built up during the onset phase, it steadily increases during scratching, as the frontal pileup increases in size.

\item The normal hardness assumes similar values as during indentation. The magnitude of the transverse hardness depends critically on whether the frontal pileup is included in the area determination or not; the hardness decreases significantly if the pileup is included since it offers less resistance to the scratching tip than the subsurface material. If only the submersed tip area is taken into account for hardness determination, the transverse hardness is of similar size as the normal hardness. 

\item We quantify the plastic activity by the volume of active zones, in which atoms were subject to a strain larger than a pre-defined cutoff value. While, evidently,  the absolute values of the volumes depend on the cutoff value, the production rates do not. They allow us to define a plastic efficiency, \qeq{e_eff}, as the size of the plastic zone relative to the  size of the excavated groove. This quantity thus relates internal damage (plasticity) to external damage (abrasive wear). Its value is similar in indentation and scratching, and shows only little temperature dependence up to the glass transition temperature. The strong  increase of the efficiency when  scratching at the glass temperature suggests that the indentation process reduces the activation barriers for structural rearrangement in a large region of the sample, thus allowing more structural relaxation processes which contribute to faster increase of the volume of active zones during scratching.

\item These active zones correspond to a set of STZs. An organization in (planar) shear bands is not recognizable for scratching depths $<30$ \AA. Only for larger depths the formation of shear bands is initiated. However, the formed shear bands show no effects, other than proportional, in the macroscopic quantities measured here. 

\item The glass material becomes significantly softer with increasing temperature, in that its (normal and transverse) hardness  decreases. The friction coefficient is remarkably unaffected by temperature. Below the glass transition temperature, we cannot identify a significant change in the friction coefficient. However, above the glass transition temperature, the friction coefficient rises.

\item A peculiar behavior becomes apparent at or slightly above the glass transition temperature in that the generation of active zones increases whereas the height of the pileup decreases.

\item We find that at high temperatures, close to the glass transition temperature, a decrease in hardness correlates with an increase of the friction coefficient, higher generation rate of the plastic zones and lower elastic and visco-elastic response. This suggests that the increase in the friction coefficient is caused by enhanced activation of atomic rearrangements at higher temperatures.

\end{enumerate}

\appendix

\section{Contact area determination}\label{appendix2}

In this appendix, we 
 introduce a new method to determine the contact area in nanoscratching. First, we briefly summarize the most common methods used for spherical tips.

\begin{figure}[h]
  \begin{center}
    \includegraphics[width=0.45\textwidth]{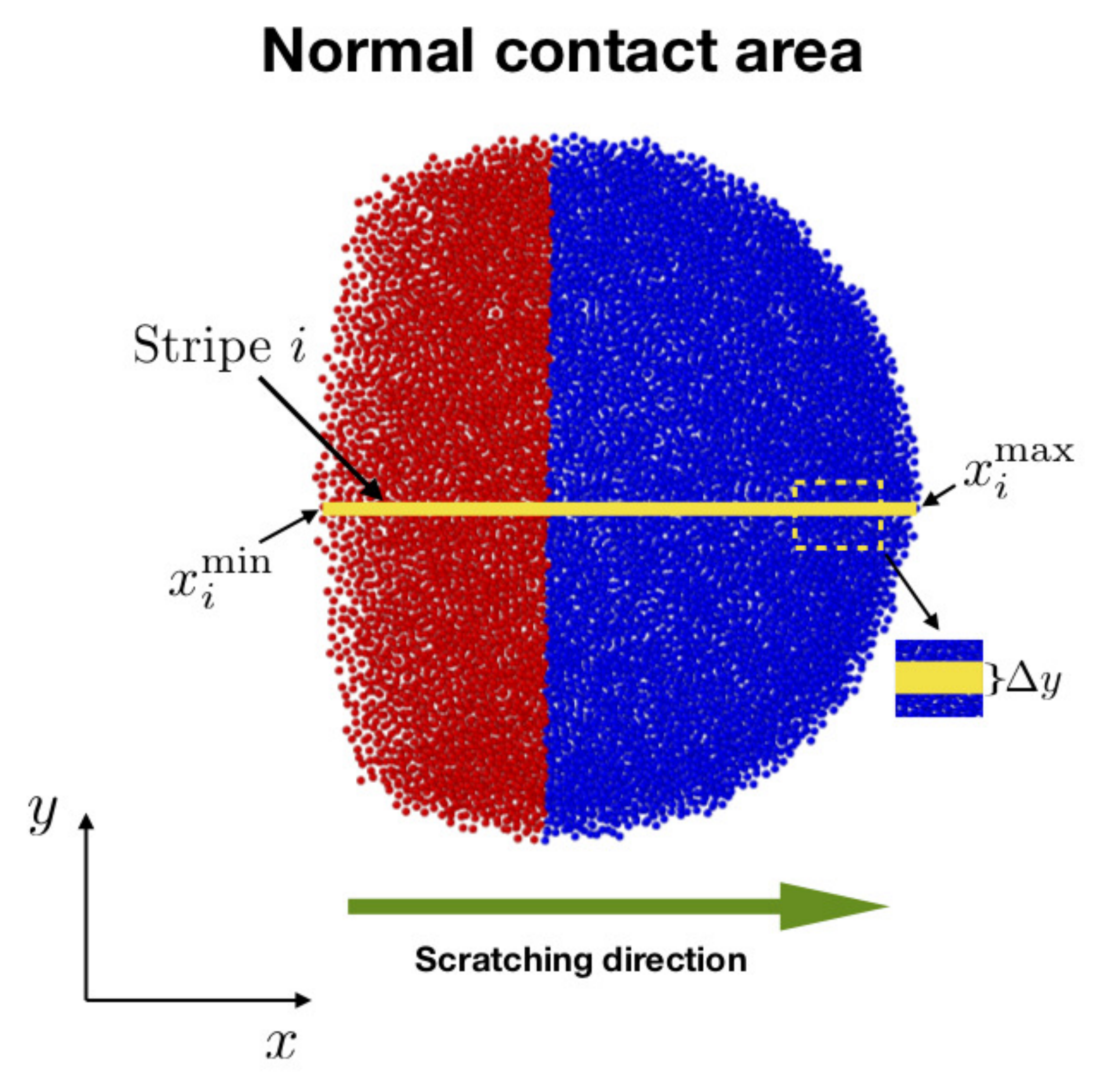}
     \end{center}
  \caption{Projected normal area of atoms contacting the indenter in a scratch simulation. The front (rear) part of the tip is marked by different colors. The contact area is calculated by cutting the geometrical projection in stripes of width $\Delta y$ and length $x_i^{\rm max} -x_i^{\rm min}$. 
  }
  \label{fig:A1}
\end{figure}

For indentation, the normal contact area is usually determined by geometrical methods. The most straightforward method is the approximation of the projected area by a circle. This is used  both in experiments and simulation. A more accurate method used in MD simulations is the elliptical approximation~\cite{ZUH10jap}. The elliptical method determines the area of an ellipse by using the minimum and maximum position of the atoms, both in x and y directions, in the following way
\begin{equation}
A_{\text{elliptical}}=\frac{\pi}{4}(x_{\text{max}}-x_{\text{min}})(y_{\text{max}}-y_{\text{min}}).
\end{equation}
The atoms considered to be in contact with the indenter are contained within a shell separated by a distance $r_c$ from the indenter. In the following methods, it is also necessary to select a shell around the indenter to determine the contact atoms. The bigger $r_c$, the bigger is the estimated value of the area.

In nanoscratching, the area determination is not so straightforward since it is  irregularly shaped (see Fig.~\ref{fig:A1}). In Refs.~\cite{ZUH10jap,GBKU15,AU16}, it was proposed to sum the projected areas of individual atoms in contact with the indenter in order to calculate the total contact area, which is referred to as the atomistic method. Each atom has an area $\pi \sigma^2$ projected according to the angle $\alpha_i$ formed from the center of the indenter to the position of the contact atom $i$. The normal and transverse contact areas are thus given by 
 \begin{align}
 A_{\text{atom}}^N & =\pi \sigma^2 \sum_{i \in {\text {contact}}} \cos{\alpha_i} , \label{atomistic} \\
 A_{\text{atom}}^T & =\pi \sigma^2 \sum_{i \in {\text {contact}}} \sin{\alpha_i} \cos{\theta_i} .
 \end{align}
 Note that for the calculation of  the tangential area, also the azimuthal angle, $\theta_i$, under which atom $i$ is seen with respect to the scratch direction, had to be introduced~\cite{ABKU17}.

The atomistic method has been implemented with success in several simulations of nanoscratching in crystals~\cite{ZUH10jap,GBKU15,AU16}. However, for substrates containing more than one atom species, the introduction of a second or more values of $\sigma$ complicates the calculation of the contact areas using this method. 

Here we propose to calculate the contact area by a simple one-dimensional integration scheme. To this end, we divide up  the projected area by stripes (along the scratching direction) of width $\Delta y$. The length of each stripe is given by the difference between the atom at the maximum position and the atom at the minimum position in the direction of scratching, both located within the stripe (see Fig.~\ref{fig:A1}). In the case of the normal projection, the contact area is given by
\begin{equation}
A_{\text{Int}}^N=\sum_{i=1} ^ {N_{\text{total}}} (x_i^{\rm max} -x_i^{\rm min}) \Delta y,
\end{equation}
where $N_{\text{total}}$ is the total number of  stripes. The transverse projection is  given by
\begin{equation}
A_{\text{Int}}^T=\sum_{i=1} ^ {N_{\text{total}}} (z_i^{\rm max}-z_i^{\rm min}) \Delta y.
\end{equation}
For both cases, normal and transverse calculations, we select the same stripe width  $\Delta y$. In the present work, we select $\Delta y=1.6$ \AA, which is about the value of the atomic radius of the larger of two atom species (Zr)~\cite{MSK07}, and $r_c=2.77$ \AA, which corresponds to the first maximum of the radial distribution function~\cite{MSK07}.

\begin{figure}[b]
  \begin{center}
    \includegraphics[width=0.45\textwidth]{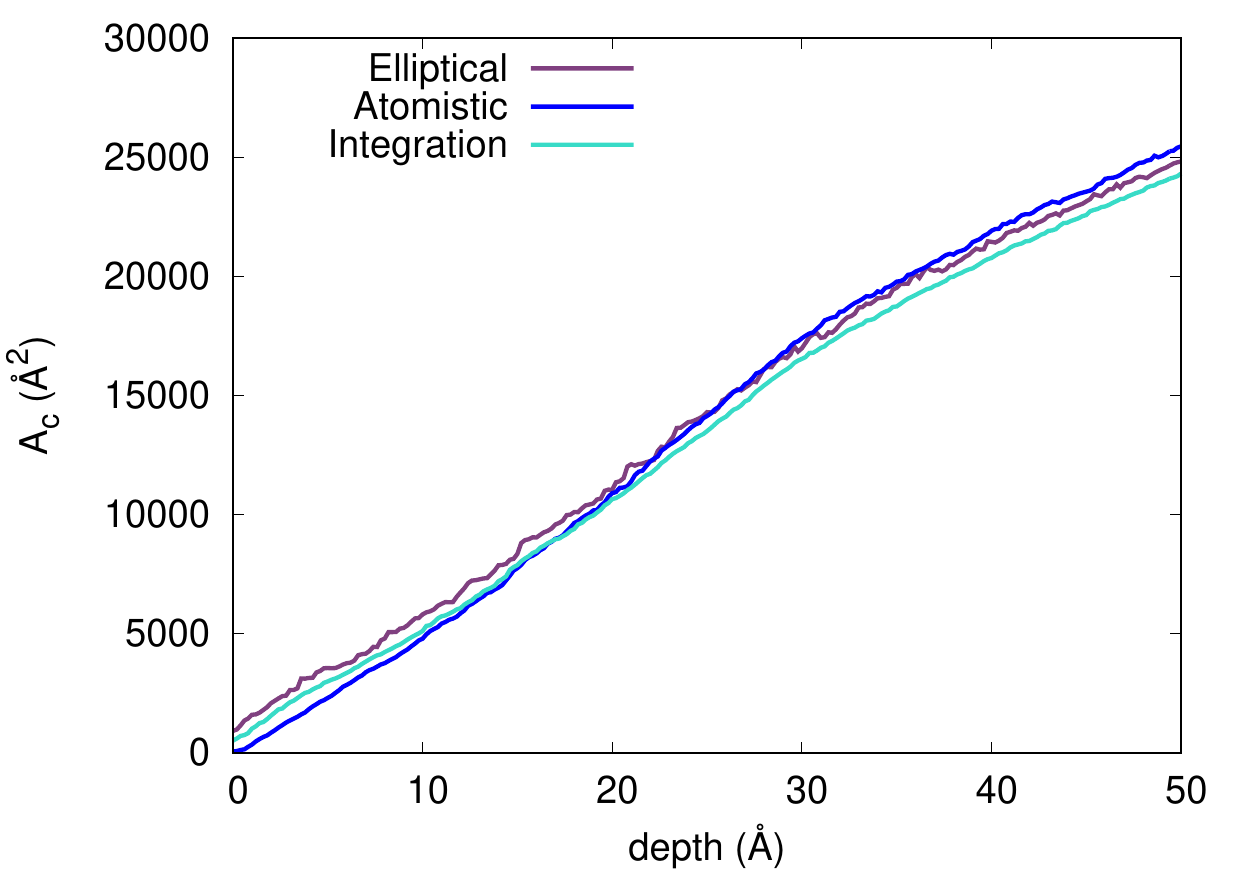}
     \end{center}
  \caption{Evolution of the normal contact area with indentation depth as  determined with the methods discussed in Appendix \ref{appendix2}.
  }
  \label{fig:A2}
\end{figure}

For the case of nanoindentation, we compare the results of  the elliptical, atomistic and the integration method in Fig.~\ref{fig:A2}.  
In this figure we use $r_c=2.77$ \AA\  for the elliptical and the integration method. Since the atomistic method measures the real projection of the atoms around the indenter, we restrict $r_c$ to strictly the first nighest neighbors shell by using half of the peak from the radial distribution function, i.e., $r_c=1.385$~\AA. Also, for the atomistic method, we use $\sigma=1.5$ \AA\ (see Eq. (\ref{atomistic})), which corresponds to the weighted radius by the chemical composition percentage. We can observe that all methods follow each other closely.

\begin{acknowledgments}

We acknowledge support by the Deutsche Forschungsgemeinschaft via the SFB/TRR 173. 
Access to the computational resources provided by the compute cluster `Elwetritsch' of the University of Kai\-sers\-lau\-tern is appreciated. 

\end{acknowledgments}

\bibliography{../../string,../../Glasses,../../PUBL,../../all}
%\bibliography{/Users/urbassek/Documents/bib/base/string,/Users/urbassek/Documents/bib/base/all,/Users/urbassek/Documents/bib/base/publ}

%\newpage \clearpage

\newpage \clearpage

\end{document}